\documentclass[lettersize,journal]{IEEEtran}
\usepackage{array}
\usepackage[caption=false,font=normalsize,labelfont=sf,textfont=sf]{subfig}
\usepackage{textcomp}
\PassOptionsToPackage{hyphens}{url}\usepackage{hyperref}
\usepackage{verbatim}
\usepackage{cite}
\usepackage{amsmath,amssymb,amsthm,amsfonts,bbm}
\usepackage{graphicx}
\usepackage[dvipsnames]{xcolor}
\usepackage{multirow}
\usepackage{subcaption}
\usepackage{caption}
\usepackage{dsfont}
\usepackage{soul}
\usepackage{subfiles}
\usepackage{comment}
\usepackage{booktabs}
\usepackage{epstopdf} 
\usepackage[table,xcdraw]{xcolor}
\usepackage{pifont}
\usepackage[flushleft]{threeparttable}

\def\BibTeX{{\rm B\kern-.05em{\sc i\kern-.025em b}\kern-.08em
    T\kern-.1667em\lower.7ex\hbox{E}\kern-.125emX}}

\usepackage{tikz}
\usetikzlibrary{fit,calc}

\colorlet{pink}{red!40}
\colorlet{blue}{cyan!50}

\usepackage[acronym]{glossaries}
\makeglossaries
\newacronym{ieee}{IEEE}{Institute of Electrical and Electronics Engineers}
\newacronym{vr}{VR}{Virtual Reality}
\newacronym{ar}{AR}{Augmented Reality}
\newacronym{obss}{OBSS}{Overlapping Basic Service Set}
\newacronym{iot}{IoT}{Internet of Things}
\newacronym[plural=APs]{ap}{AP}{Access Point}
\newacronym{csmaca}{CSMA/CA}{Carrier Sense Multiple Access with Collision Avoidance}
\newacronym{lbt}{LBT}{Listen-Before-Talk}
\newacronym[plural=BSSs]{bss}{BSS}{Basic Service Set}
\newacronym{mlo}{MLO}{Multi-Link Operation}
\newacronym{sr}{SR}{Spatial Reuse}
\newacronym{obsspd}{OBSS/PD}{Overlapping Basic Service Set/Packet Detect}
\newacronym[plural=STAs]{sta}{STA}{Station}
\newacronym{snr}{SNR}{Signal-to-Noise-Ratio}
\newacronym{tgbn}{TGbn}{Task Group bn}
\newacronym{cosr}{Co-SR}{Coordinated Spatial Reuse}
\newacronym{cobf}{Co-BF}{Coordinated Beamforming}
\newacronym{cortwt}{Co-RTWT}{Coordinated R-TWT}
\newacronym{cotdma}{Co-TDMA}{Coordinated Time Division Multiple Access}
\newacronym{jt}{JT}{Joint Transmission}
\newacronym{coofdma}{Co-OFDMA}{Coordinated OFDMA}
\newacronym{mapc}{MAPC}{Multi-Access Point Coordination}
\newacronym{txop}{TXOP}{Transmission Opportunity}
\newacronym{ecsr}{ECSR}{Enhanced Coordinated Spatial Reuse}
\newacronym{rssi}{RSSI}{Received Signal Strength Indicator}
\newacronym{mcs}{MCS}{Modulation and Coding Scheme}
\newacronym{dcf}{DCF}{Distributed Coordination Function}
\newacronym{sinr}{SINR}{Signal-to-Interference-plus-Noise Ratio}
\newacronym{rrm}{RRM}{Radio Resource Management}
\newacronym{rts}{RTS}{Request-to-Send}
\newacronym{cts}{CTS}{Clear-to-Send}
\newacronym{arq}{ARQ}{Automatic Repeat Request}
\newacronym{ampdu}{A-MPDU}{Aggregate MAC Protocol Data Unit}
\newacronym{ai}{AI}{Artificial Intelligence}
\newacronym{ml}{ML}{Machine Learning}
\newacronym{phy}{PHY}{Physical}
\newacronym{mac}{MAC}{Medium Access Control}
\newacronym{ac}{AC}{Access Category}
\newacronym{uhr}{UHR}{Ultra High Reliability}
\newacronym{sifs}{SIFS}{Short Interframe Space}
\newacronym{difs}{DIFS}{DCF Interframe Space}
\newacronym{aifs}{AIFS}{Arbitration Interframe Space}
\newacronym{wds}{WDS}{Wireless Distribution System}
\newacronym{ssid}{SSID}{Service Set Identifier}
\newacronym{ess}{ESS}{Extended Service Set}
\newacronym{mbss}{MBSS}{Mesh Basic Service Set}
\newacronym{wlc}{WLC}{Wireless LAN Controller}
\newacronym{wlan}{WLAN}{Wireless Local Area Network}
\newacronym{capwap}{CAPWAP}{Control And Provisioning of Wireless Access Points}
\newacronym{ietf}{IETF}{Internet Engineering Task Force}
\newacronym{rfc}{RFC}{Request for Comments}
\newacronym{isp}{ISP}{Internet Service Provider}
\newacronym{cpe}{CPE}{Customer-Premises Equipment}
\newacronym{usp}{USP}{User Services Platform}
\newacronym{mqtt}{MQTT}{Message Queuing Telemetry Transport}
\newacronym{bssid}{BSSID}{Basic Service Set Identifier}
\newacronym{tpc}{TPC}{Transmit Power Control}
\newacronym{cocr}{Co-CR}{Coordinated Channel Recommendation}
\newacronym{pasn}{PASN}{Pre-Association Security Negotiation}
\newacronym{ltf}{LTF}{Long Training Field}
\newacronym{gi}{GI}{Guard Interval}
\newacronym{twt}{TWT}{Target Wake Time}
\newacronym{sp}{SP}{Service Period}
\newacronym{rtwt}{R-TWT}{Restricted Target Wake Time}
\newacronym{aid}{AID}{Association ID}
\newacronym{pmksa}{PMKSA}{Pairwise Master Key Security Association}
\newacronym{ptksa}{PTKSA}{Pairwise Transient Key Security Association}
\newacronym{rsne}{RSNE}{Robust Security Network Element}
\newacronym{rsnxe}{RSNXE}{RSN eXtension Element}
\newacronym{mic}{MIC}{Message Integrity Check}
\newacronym{back}{BACK}{Block Acknowledgment}
\newacronym{ppdu}{PPDU}{Physical Layer Protocol Data Unit}
\newacronym{csi}{CSI}{Channel State Information}
\newacronym{ndp}{NDP}{Null Data Packet}
\newacronym{bsrp}{BSRP}{Buffer Status Report Poll}
\newacronym{ntb}{NTB}{Non-Trigger Based}
\newacronym{icf}{ICF}{Initial Control Frame}
\newacronym{icr}{ICR}{Initial Control Response}
\newacronym{dps}{DPS}{Dynamic Power Save}
\newacronym{emlsr}{EMLSR}{Enhanced Multi-Link Single Radio}
\newacronym{ofdm}{OFDM}{Orthogonal Frequency-Division Multiplexing}
\newacronym{eht}{EHT}{Extremely High Throughput}
\newacronym{ndpa}{NDPA}{NDP Announcement}
\newacronym{tb}{TB}{Trigger-Based}
\newacronym{bfrp}{BFRP}{Beamforming Report Poll}
\newacronym{ofdma}{OFDMA}{Orthogonal Frequency-Division Multiple Access}
\newacronym{mu}{MU}{Multi-User}
\newacronym{txs}{TXS}{Triggered TXOP Sharing}
\newacronym{mpdu}{MPDU}{MAC Protocol Data Unit}
\newacronym{tsf}{TSF}{Time Synchronization Function}
\newacronym{p2p}{P2P}{Peer-to-Peer}
\newacronym{txspg}{TXSPG}{TXOP Sharing for P2P Groups}
\newacronym{scs}{SCS}{Stream Classification Service}
\newacronym{qos}{QoS}{Quality of Service}
\newacronym{ru}{RU}{Resource Unit}
\newacronym{mimo}{MIMO}{Multiple Input Multiple Output}
\newacronym{lli}{LLI}{Low Latency Indicator}
\newacronym{cdf}{CDF}{Cumulative Distribution Function}
\newacronym{npca}{NPCA}{Non-Primary Channel Access}
\newacronym{conpca}{Co-NPCA}{Coordinated NPCA}
\newacronym{cfo}{CFO}{Carrier Frequency Offset}
\newacronym{ula}{ULA}{Uniform Linear Array}
\newacronym{xr}{XR}{eXtended Reality}
\newacronym{ht}{HT}{High Throughput}
\newacronym{lsig}{L-SIG}{Legacy Signal Field}
\newacronym{bar}{BAR}{Block ACK Request}
\newacronym{ack}{ACK}{Acknowledgement}
\newacronym{he}{HE}{High Efficiency}
\newacronym{edca}{EDCA}{Enhanced Distributed Channel Access}
\newacronym{mld}{MLD}{Multi-Link Device}
\newacronym{duo}{DUO}{Dynamic Unavailability Operation}
\newacronym{tcp}{TCP}{Transmission Control Protocol}
\newacronym{zf}{ZF}{Zero-Forcing}
\newacronym{udp}{UDP}{User Datagram Protocol}

\begin{document}

\bstctlcite{IEEEexample:BSTcontrol} 

\title{A Tutorial on IEEE 802.11bn Multi-AP Coordination for Wi-Fi 8: From Standardization to Performance Evaluation}

\author{Francesc Wilhelmi, Boris Bellalta, Giovanni Geraci, Lorenzo Galati-Giordano, Francesca Meneghello,\\ Aleksandra Kijanka, Iñaki Val, David López-Pérez 
\thanks{Francesc Wilhelmi, Boris Bellalta, and Aleksandra Kijanka are with Universitat Pompeu Fabra (Spain). 
Giovanni Geraci is with Nokia (Spain) and Universitat Pompeu Fabra (Spain). 
Lorenzo Galati-Giordano is with Nokia Bell Labs (Germany). 
Francesca Meneghello is with Northeastern University (USA). 
Iñaki Val is with MaxLinear (Spain). 
David López-Pérez is with Universitat Politècnica de València (Spain). 
}
}

\markboth{Journal of \LaTeX\ Class Files,~Vol.~14, No.~8, August~2021}%
{Shell \MakeLowercase{\textit{et al.}}: A Sample Article Using IEEEtran.cls for IEEE Journals}

\IEEEpubid{0000--0000/00\$00.00~\copyright~2021 IEEE}

\markboth{Journal of \LaTeX\ Class Files,~Vol.~14, No.~8, August~2021}%
{Shell \MakeLowercase{\textit{et al.}}: A Sample Article Using IEEEtran.cls for IEEE Journals}

\IEEEpubid{0000--0000/00\$00.00~\copyright~2021 IEEE}

\maketitle

\thispagestyle{plain}
\pagestyle{plain}

\begin{abstract}
The IEEE 802.11bn amendment defines significant modifications to the standard by establishing Ultra High Reliability (UHR) targets in Wireless Local Area Networks (WLANs). This is expected to deliver substantial enhancements over previous standards, including modes of operation that increase throughput, reduce the 95th percentile of the latency distribution, and decrease MAC Protocol Data Unit (MPDU) loss (all by at least 25\%) compared to Extremely High Throughput (EHT) operations defined in the 802.11be amendment. A fundamental innovation for achieving these ambitious goals is the introduction of Multi-Access Point Coordination (MAPC), an unprecedented feature whereby APs will be able to coordinate among themselves to enhance spectrum utilization and advance towards reliability. This paper provides a comprehensive overview and analysis of this key framework. We begin by reviewing existing AP coordination solutions that precede the 802.11bn standard, which serve as a foundation for understanding the transition to the current framework. We then describe the technical 802.11bn MAPC framework as defined by the task group. A detailed overview of each candidate MAPC feature is provided, contextualized with the relevant state-of-the-art. Furthermore, we introduce \texttt{Kom8ndor}, an open-source Wi-Fi~8 simulation tool, to evaluate these candidate MAPC features and showcase their potential to achieve UHR goals. Finally, we outline the future of MAPC beyond 802.11bn, exploring promising directions such as coordination schemes beyond 802.11bn (e.g., Joint Transmission (JT)) and new ideas.
\end{abstract}


\section{Introduction}
\label{sec:introduction}

\glsunset{mapc} 

\IEEEPARstart{M}{ulti-Access Point Coordination} (MAPC) emerges as one of the central pillars of the upcoming \gls{ieee} 802.11bn (11bn) amendment, which aims to deliver ``\gls{uhr}'' capabilities, a performance milestone not previously targeted in Wi-Fi~\cite{geraci2025wi,reshef2022future, galati2024will}. By defining new \gls{mac}/\gls{phy} functionalities and signaling between \gls{ap} devices from different \glspl{bss}, \mbox{Wi-Fi} 8 is expected to enhance channel access operations and determinism, improving performance and efficiency. Five main mechanisms operating on a microsecond scale (e.g., at the \gls{txop} level) are currently being developed for 11bn (see Table~\ref{tab:mapc_features})---namely, \gls{cosr}, \gls{cobf}, \gls{cotdma}, \gls{cortwt}, and \gls{cocr}---but \gls{mapc} is expected to evolve further in future Wi-Fi amendments, potentially including features already discussed within 802.11, such as \gls{jt} or \gls{coofdma}.

\gls{mapc} is unprecedented in Wi-Fi, so only a small set of literature works is available. An early survey on \gls{mapc} can be found in~\cite{verma2023survey}, while more recent works provide an overview of Wi-Fi 8 functionalities and include aspects of \gls{mapc}~\cite{val2025wi, geraci2025wi, charfi2025wi}. Wi-Fi \gls{mapc} has been discussed since the days of 802.11be (11be). This led to the proliferation of academic studies such as~\cite{zhang2023ieee}, which investigated the benefits of \gls{mlo} coordination (a feature not included in 11bn), or \cite{imputato2022multi}, which anticipated \gls{cotdma} (included in 11bn). However, due to their early publication, these works do not provide sufficient details about the current coordination architecture and mechanisms defined in 11bn. 

\begin{table*}[ht!]
\caption{Summary of the main MAPC schemes defined in the IEEE 802.11bn.}
\label{tab:mapc_features}
\resizebox{\textwidth}{!}{%
\begin{tabular}{@{}cllc@{}}
\toprule
\textbf{\begin{tabular}[c]{@{}c@{}}MAPC \\scheme\end{tabular}} & \multicolumn{1}{c}{\textbf{Primary objective}} & \multicolumn{1}{c}{\textbf{Core mechanism}} & \textbf{\begin{tabular}[c]{@{}c@{}}Layer of \\operation\end{tabular}} \\ \midrule
\gls{cobf} & Enhanced medium utilization & Multi-AP beamforming with null steering for concurrent transmissions & \gls{phy} \\
\gls{cosr} & Enhanced medium utilization & Coordinated concurrent downlink transmission through power control & \gls{phy}/\gls{mac} \\
\gls{cotdma} & Efficient resource sharing & Subdivision of a TXOP and inter-AP scheduling  & \gls{mac} \\
\gls{cortwt} & Protection for scheduled traffic & Coordination of R-TWT schedules between OBSS & \gls{mac} \\ 
\gls{cocr} & P2P communication enhancement & Recommended channels list for P2P transmissions & \gls{mac} \\ \bottomrule
\end{tabular}%
}
\end{table*}

This paper provides a detailed description of 11bn's \gls{mapc}, blending its technical standardization---architectural design, operation, signaling, and frame formatting---with an illustrative guideline and performance evaluation, serving as a comprehensive reference for both researchers and practitioners. Furthermore, it explores future directions for \gls{mapc} beyond 11bn by examining past discussions, current trends, and relevant emerging technologies within 802.11. To the best of our knowledge, this is the first work to overview the technical details of 11bn Draft 1.4 (D1.4)~\cite{11bn_draft}, which represents an important milestone in the development of the amendment. In addition, it is the first work to compare multiple \gls{mapc} schemes with the same tool and a consistent methodology. Specifically, the main contributions of this paper are as follows:
\begin{itemize}
    \item We overview existing solutions for \gls{ap} coordination prior to 11bn (Section~\ref{sec:wifi_history}).
    \item We describe the technical 11bn \gls{mapc} framework, as defined by the \gls{tgbn} (Section~\ref{sec:mapc_framework}).
    \item We provide a detailed overview of each candidate \gls{mapc} feature for 11bn. Each feature is described along with relevant state-of-the-art (Section~\ref{sec:mapc_features}).
    \item Using \texttt{Kom8ndor}, a unified, open-source Wi-Fi 8 simulation tool ~\cite{barrachina2019komondor}, we evaluate three candidate \gls{mapc} features (\gls{cosr}, \gls{cobf}, and \gls{cotdma}) in carefully designed scenarios to showcase their main tradeoffs (Section~\ref{sec:simulations}). These schemes operate at the \gls{txop} level and directly shape the exchange of packets, hence their measurable performance is immediate and easily comparable to a non-coordinated baseline. In contrast, \gls{cortwt} and \gls{cocr} operate at a higher level of abstraction, coordinating schedules and channel assignments without directly impacting individual transmissions. The evaluation is purposefully didactic and serves to understand when and why each mechanism is beneficial.    
    \item We discuss the future of \gls{mapc} beyond Wi-Fi~8, including features like \gls{jt} and \gls{coofdma} and other relevant directions in standardization (Section~\ref{sec:future_mapc}).
\end{itemize}

\section{From Mesh to Coordination: A Glimpse of the Evolution of Wi-Fi Towards MAPC}
\label{sec:wifi_history}

The cooperation among \glspl{ap} has been a relevant topic since the beginning of 802.11 due to the inherent issues of decentralized operation, and has evolved from centralized, proprietary enterprise architectures to the framework standardized in 802.11bn. In this section, we overview the role of 802.11 amendments before 802.11bn that provide inter-\gls{ap} coordination, the emergence of centralized controllers, and advanced mesh systems.

\subsection{IEEE 802.11 AP Coordination Before 802.11bn} 

\subsubsection{Wireless Distribution System (WDS)} 

Since the beginning of IEEE 802.11 (1999), a feature called \gls{wds} has enabled wireless connections between \glspl{ap} in the same 802.11 network, expanding it without needing wired links and providing mesh-like capabilities~\cite{el2000implementation}. In particular, in \gls{wds}, different \glspl{ap} with the same \gls{ssid} are used to expand the coverage area of a given \gls{bss}, thus forming a wireless \gls{ess}. \gls{wds} APs can be assigned different roles, including \emph{main \gls{ap}} (it gives access to the non-802.11 network through, e.g., Ethernet), \emph{relay \gls{ap}} (forwards traffic between main and remote \glspl{ap}), and \emph{remote \gls{ap}} (serves the \glspl{sta} and participates in the mesh).
Based on these roles, two different \gls{ap} operational modes are defined: \emph{i)} \emph{Wireless bridging}, in which the \gls{ap} operates as a relay \gls{ap} forwarding traffic between network segments without serving \glspl{sta} directly; and \emph{ii)} \emph{Wireless repeating}, in which the \gls{ap} operates as a remote \gls{ap}, simultaneously maintaining upstream connectivity to the main or relay \gls{ap} and serving associated \glspl{sta}. To realize packet forwarding, an \gls{ap} transmits 802.11 \gls{mac} frames with multiple addresses, which are intended to identify the destination, source, transmitter, and addressee. When it comes to packet encryption, all the \glspl{ap} in a \gls{wds} must apply the same method (e.g., none, WEP, WPA, WPA2) and share the same encryption keys. The \gls{wds} technology, while providing mesh capabilities since the very beginning of 802.11, is now deprecated. Moreover, as the standard left much room for interpretation, \gls{wds} devices implemented by different vendors are often non-interoperable.

\subsubsection{IEEE 802.11s (Mesh)}

The 802.11s is an extension of 802.11 to provide mesh networking capabilities that was initiated in 2003 (with the creation of the Mesh Study Group) but was not included in the 802.11 specifications until 2012~\cite{carrano2010ieee, wang2008ieee}. The 802.11s allows for mesh networking by establishing a \gls{mbss}, i.e., a set of interconnected multi-hop \glspl{bss}.\footnote{In \gls{mbss}, a given mesh node may also be an AP deploying its own \gls{bss}. In an ESS, in contrast, multiple \gls{ap} (connected through a backhaul) are used to extend the coverage of a single \gls{bss}, so they appear to be only one from the point of view of the clients.} The operation of the 802.11s is anchored in Layer 2, so that it provides a bridged Layer 2 network on which any Layer 3 could operate. The 802.11 covers the definition and establishment of mesh nodes (referred to as \textit{Mesh \glspl{sta}}), the establishment and management of meshed routes, and the transportation of meshed messages. In addition to that, the 802.11s includes mechanisms such as the Mesh Coordination Function (MCF) and Mesh Coordinated Channel Access (MCCA) to provide deterministic channel access, congestion control, and power saving.

\subsubsection{Other relevant IEEE 802.11 amendments}

Several 802.11 amendments have contributed to empowering \mbox{Wi-Fi} networks with certain multi-\gls{ap} capabilities, leading to improved resource allocation. Some relevant examples are 802.11k (11k), 802.11v (11v), and 802.11r (11r), which are extensions of the 802.11 standard that collectively provide a framework for network-assisted \gls{sta} mobility, hence forming the basis for modern multi-\gls{ap} systems such as Wi-Fi EasyMesh (see Section~\ref{sec:easymesh}). More in detail:
\begin{itemize}
    \item 11k (Radio Resource Measurement) enables the exchange of radio resource measurements between the \gls{ap} and its client \glspl{sta} in the form of \textit{Neighbor Reports}~\cite{adame2021channel}. Such reports include a list of nearby \glspl{ap} and their main characteristics (e.g., \gls{bssid}, operating channel, capabilities), which can be acquired by overhearing management frames like Beacons. Although intended for roaming, the 11k allows Wi-Fi devices to better understand their operating environment, which is useful to make smart decisions at different levels (e.g., select a better channel).
    \item 11v (\gls{bss} Transition Management) builds on top of 11k to provide a management framework for network-assisted roaming, where an \gls{ap} can suggest \glspl{sta} to associate with other better \glspl{ap} (which can be identified considering the current load and \gls{rssi} of neighboring \glspl{ap})~\cite{adame2021channel}.    
    \item 11r (Fast \gls{bss} Transition) focuses on reducing the latency of roaming operations by enhancing the security handshake during a new authentication and association~\cite{bangolae2006performance}. In particular, 11r defines a mechanism to pre-establish security keys between an \gls{sta} and a potential \gls{ap} to roam. This is achieved by defining a \textit{mobility domain}, whereby multiple \glspl{ap} can share security keys.
\end{itemize}

\subsection{Centralized Controller Solutions} 

In spite of not working at the \gls{txop} level, the earliest and most robust form of \gls{mapc} emerged in the enterprise sector. The controller-based paradigm decouples the data plane (in the AP) from the control plane (in a centralized \gls{wlc}) and has been typically used to provide centralized \gls{rrm} (including dynamic channel assignment and transmit power control), load balancing (to ensure clients are properly distributed among \glspl{ap}), or mobility (proactively managing client handovers). Depending on the type of network (enterprise vs. residential controllers), different solutions have been applied:
\begin{itemize}
    \item In enterprise/industrial solutions, a widely adopted technology to enable \glspl{wlc} is \gls{capwap}~\cite{levanti2007capwap}, a transport layer protocol defined in 2009 by \gls{ietf} in \gls{rfc} 5415 (the binding with 802.11 was specified in \gls{rfc} 5416~\cite{calhoun2009rfc}) to control wireless-agnostic termination points from a controller entity that uses 802.3 encapsulation. \gls{capwap} defines \emph{Access Controller} and \emph{Wireless Termination Point} entities, where the former not only controls the operation of the latter (through configurations and firmware updates) but also takes care of user authentication and management, including admission control. The functions defined by \gls{capwap} include Distribution/Integration service, Beacon generation, Probe Response generation, Power management and Packet buffering, Fragmentation/Defragmentation, Association/Disassociation/Reassociation, QoS (classifying, scheduling, queuing), 802.1X/EAP, RSNA key management, and 802.11 encryption/decryption. Accordingly, proprietary solutions based on \gls{capwap} can implement \gls{rrm} functionalities and gather 802.11 measurements (e.g., success/failures per type of frame).
    \item In residential Wi-Fi controllers, network management protocols like \gls{usp} (TR-369) are often adopted by \gls{isp} to manage \gls{cpe} such as Wi-Fi \glspl{ap}~\cite{cheng2024wi}. Such protocols allow for managing, controlling, and monitoring 802.11 \glspl{ap} from a central controller. TR-369 uses persistent connection protocols like WebSockets and \gls{mqtt} for real-time communication with the managed \glspl{ap}, thus allowing for configuring and monitoring all the connected devices. To do so, TR-369 uses the data model from TR-181, which structures 802.11 information into \texttt{\gls{ssid}} (e.g., name of the \gls{ssid}), \texttt{security} (e.g., security mode enabled, key passphrase), and \texttt{radio} (e.g., whether the radio is on or not, statistics of the radio such as bytes transmitted/sent), thus allowing to interact with both \glspl{ap} and clients.
\end{itemize}

\subsection{Wi-Fi EasyMesh}
\label{sec:easymesh}

Wi-Fi CERTIFIED EasyMesh~\cite{wifi_easymesh} (first released in 2018) is the certification program for the Multi-Access Point (MAP) technical specification,\footnote{Note that Wi-Fi CERTIFIED EasyMesh differs from the mesh certification of 802.11s.} whereby multiple Wi-Fi \glspl{ap} can be connected and configured together. Wi-Fi EasyMesh emerged because of the increased need for home Wi-Fi solutions (for extending in-home connectivity) and in response to the lack of interoperable solutions for operators to manage home Wi-Fi \glspl{ap}. Building upon the 802.11k/v/r amendments and the IEEE 1905.1 standard for home networking (which defines the setup, configuration, and operation of home networking devices, so that different technologies like Ethernet and Wi-Fi can be interconnected within the same network), EasyMesh defines a hierarchical architecture in which a controller oversees multiple subscribed agents deployed in other \glspl{ap}. In such an architecture, the controller is a root \gls{ap} that manages and configures the network, while the rest of the \glspl{ap} behave either as clients in the upstream (to the root) or as \glspl{ap} in the downstream (they serve both clients and other \glspl{ap} in the mesh). Because of this, the establishment of mesh links can be done by reusing existing encryption schemes such as WPA3. To allow such a dual behavior, EasyMesh defines multi-\gls{ap} devices, which contain a fronthaul \gls{ap}  to communicate with clients and an optional backhaul station that communicates with another \gls{ap}’s fronthaul (a discrete set of profiles is used to indicate the supported multi-\gls{ap} capabilities of a given device). On top of that, EasyMesh supports non-mesh aspects, which \glspl{isp} may use to optimize their customers’ Wi-Fi networks, including functionalities such as client steering (proactively moving clients to the \gls{ap} offering the best performance), band steering (guiding dual-band clients from the congested 2.4~GHz band to the higher-capacity 5~GHz or 6~GHz bands), or self-healing backhauls (dynamically re-routing \gls{ap}-to-\gls{ap} traffic in the event of a node failure or link degradation).

\section{IEEE 802.11bn MAPC Framework}
\label{sec:mapc_framework}

\begin{figure*}[ht]
    \centering
    \includegraphics[width=\textwidth]{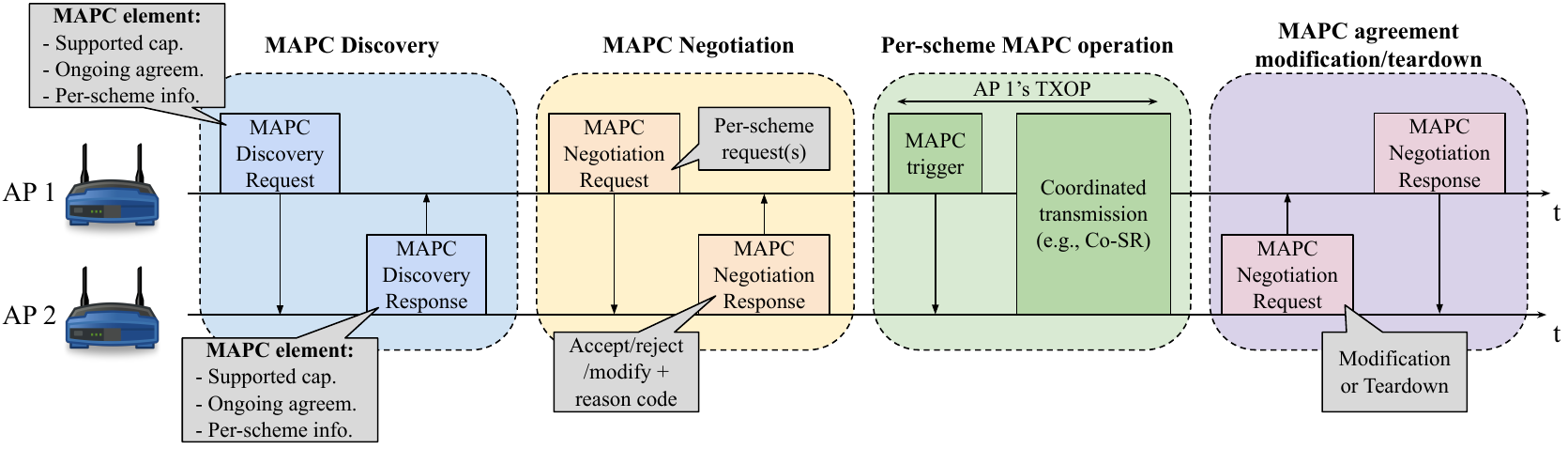}
    \caption{High-level overview of the \gls{mapc} framework. After the discovery phase, during which \glspl{ap} collect information about \gls{mapc} capabilities, a negotiation phase allows the \glspl{ap} to agree on the coordination scheme and parameters. This enables \gls{mapc} operations afterwards until teardown.}
    \label{fig:mapc_framework}
\end{figure*}

\begin{figure}[ht!]
    \centering
    \includegraphics[width=\columnwidth]{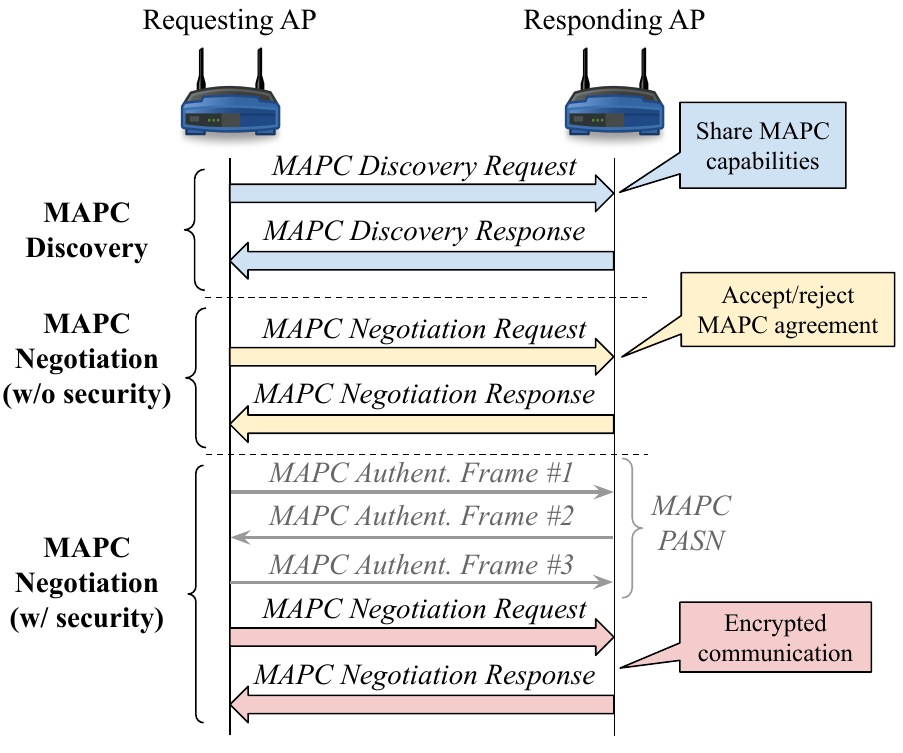}
    \caption{Messages exchanged during \gls{mapc} discovery and negotiation. These phases enable \gls{mapc} operations.}
    \label{fig:mapc_discovery_negotiation}
\end{figure}

Unlike the solutions discussed in Section~\ref{sec:wifi_history}, 11bn \gls{mapc} addresses the fundamental challenge of \textit{tight spectrum utilization efficiency} by defining coordination mechanisms that operate at the \gls{txop} level. For that, it defines a unified and flexible \gls{mapc} framework designed not only to support the coordination schemes considered within 11bn (\gls{cosr}, \gls{cobf}, \gls{cotdma}, \gls{cortwt}, and \gls{cocr}), but also future coordinated mechanisms. Specifically, 11bn defines mechanisms for \textit{\gls{mapc} Discovery}, \textit{\gls{mapc} Agreement Management} (including negotiation, update, and teardown), \textit{\gls{mapc} Authentication}, and \textit{\gls{mapc} per-scheme operation}. Depending on the selected \gls{mapc} feature, these procedures may vary in content (type of information exchanged) and shape (type of messages and periodicity). In the remainder of this section, we describe the main 11bn common procedures and signaling by which \glspl{ap} can establish and maintain coordination agreements, and leave to Section~\ref{sec:mapc_features} the in-depth description of the \textit{operation} of the different \gls{mapc} schemes. The general framework is summarized in Fig.~\ref{fig:mapc_framework} and the \gls{mapc} discovery and agreement negotiation phases are detailed in Fig.~\ref{fig:mapc_discovery_negotiation}.

\subsection{MAPC Discovery}
\label{sec:mapc_discovery}

Before establishing a coordination agreement, it is required that \glspl{ap} know the capabilities of their neighbors. For that, a discovery phase is defined, allowing \glspl{ap} to actively advertise their \gls{mapc} capabilities, which are exposed using a new \gls{mapc} Element in the body of new \gls{mapc} frames (see Appendix~\ref{app:mapc_element}). To discover candidates for coordination (see Fig.~\ref{fig:mapc_discovery_negotiation}, in blue):
\begin{enumerate}
    \item An \gls{ap} sends a new public action frame (Category 4, commonly used for standard utility actions between devices) of type \textit{\gls{mapc} Discovery Request} (see Appendix~\ref{sec:appendix_a} for details about its fields) to request \gls{mapc}-related information from neighbors. The request can be either broadcast or sent to a specific (known) \gls{ap}.
    \item The discovery request can be replied to by another \gls{ap} through another type of action frame called \textit{\gls{mapc} Discovery Response}. 
\end{enumerate}

Both request and response messages are transmitted along with the \gls{mapc} Element, which includes fields to indicate support (or lack thereof) for the different \gls{mapc} schemes. To preserve the framework's universality and extensibility in future amendments, the details of each specific scheme (e.g., the maximum number of supported sounding reports for \gls{cobf}) are informed using custom subelements.

\subsection{MAPC Agreement Management}
\label{sec:mapc_negotiation}

Once the \gls{mapc} capabilities and status from other \glspl{ap} are discovered, a requesting \gls{ap} can initiate a \gls{mapc} agreement negotiation for one or more schemes (see Fig.~\ref{fig:mapc_discovery_negotiation}, in yellow):
\begin{enumerate}
    \item First, the requesting \gls{ap} sends an \textit{\gls{mapc} Negotiation Request} (a public action frame) to another \gls{ap}. Such a message must include a request for each specific scheme, which is done by filling an \texttt{\gls{mapc} Scheme Request Set} field in each considered per-scheme subelement.
    \item A respondent \gls{ap} may respond to the requesting \gls{ap} with an \textit{\gls{mapc} Negotiation Response} message that includes a positive response or a rejection of the request (indicating the corresponding rejection reason). The acceptance/rejection of per-scheme requests is also done individually, thus the \textit{\gls{mapc} Negotiation Response} may include different subelements, one for each considered \gls{mapc} scheme. Apart from accepting or rejecting the request, the responding \gls{ap} can suggest modifying certain parameters in the agreement.
\end{enumerate}

After a successful \gls{mapc} negotiation, an \gls{mapc} agreement is established. From that moment, each \gls{ap} assigns an \gls{ap} ID to each other, which is informed in a dedicated \gls{ap} ID field (inside the MAPC Element) of the frames exchanged during the negotiation phase. Agreements can be modified or finalized by any of the cooperating \glspl{ap} by using the same \textit{\gls{mapc} Negotiation Request/Response} messages and indicating the corresponding code (e.g., \texttt{Agreement update}, \texttt{Agreement teardown}).

In 11bn, a pair of \glspl{ap} can establish up to one agreement for \gls{cobf}, \gls{cosr}, and \gls{cotdma}, and up to one agreement per \gls{rtwt} schedule for \gls{cortwt}, but multiple agreements (on different schemes) can be established between the same pair of \glspl{ap} at the same time. As an exception, in \gls{cocr}, multiple \gls{mapc} agreements can be established between the pair of \glspl{ap}.

\subsection{MAPC Authentication}
\label{sec:mapc_authentication}

The 11bn includes an optional secure way for \glspl{ap} to perform agreements, treating them as sensitive information. Whether security is used depends on the capabilities of both \glspl{ap}, which need to indicate support for \gls{mapc} security during discovery (via a \texttt{\gls{mapc} Security Supported} field in the \gls{mapc} Element) and explicitly require that all \gls{mapc} agreement negotiations are protected (by setting the \texttt{Protected Negotiations Required} field in the \gls{mapc} Element to 1). The \gls{pasn} authentication protocol is extended to the case of \gls{mapc}, which is used to establish a \gls{ptksa} that encrypts the corresponding sensitive management frames. Specifically, to carry out a secure \gls{mapc} negotiation:
\begin{enumerate}
    \item A requesting \gls{ap} sends an \textit{\gls{mapc} Authentication Request} (a new type of public action frame), which triggers the following \gls{mapc} \gls{pasn} procedure (see Fig.~\ref{fig:mapc_discovery_negotiation}, in gray):
    \begin{enumerate}
        \item The requesting \gls{ap} initiates the security negotiation and proposes the establishment of security keys. In particular, a \texttt{Security Profile} subelement is included in the \texttt{\gls{mapc} Element} to inform about the supported security protocols and parameters.   
        \item The responding \gls{ap} confirms the proposed security parameters by sending a second authentication frame, which also includes a first \gls{mic} element for integrity.
        \item The requesting \gls{ap} validates and confirms the establishment of a \gls{ptksa}, hence completing the mutual authentication procedure.     
    \end{enumerate}
    \item The same requesting \gls{ap} sends an \textit{\gls{mapc} Negotiation Request}, which in this case is a protected dual of public action frame (Category 9), but fulfills the same purpose described in Section~\ref{sec:mapc_negotiation} (in red in Fig.~\ref{fig:mapc_discovery_negotiation}).  
    \item A responder \gls{ap} replies with an encrypted \textit{MAPC Negotiation Response}  (in red in Fig.~\ref{fig:mapc_discovery_negotiation}), which is also a protected dual of public action frame.
\end{enumerate}

\begin{figure*}[ht!]
    \centering
    \subfloat[]{\includegraphics[width=0.3\textwidth]{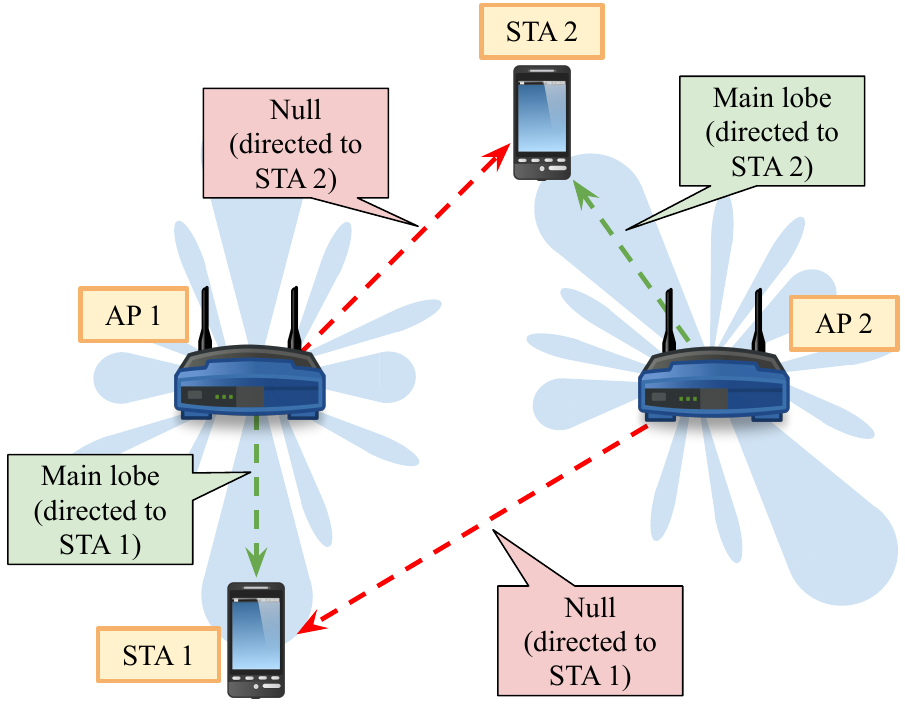}%
    \label{fig:example_beamforming}}
    \hfil
    \subfloat[]{\includegraphics[width=0.7\textwidth]{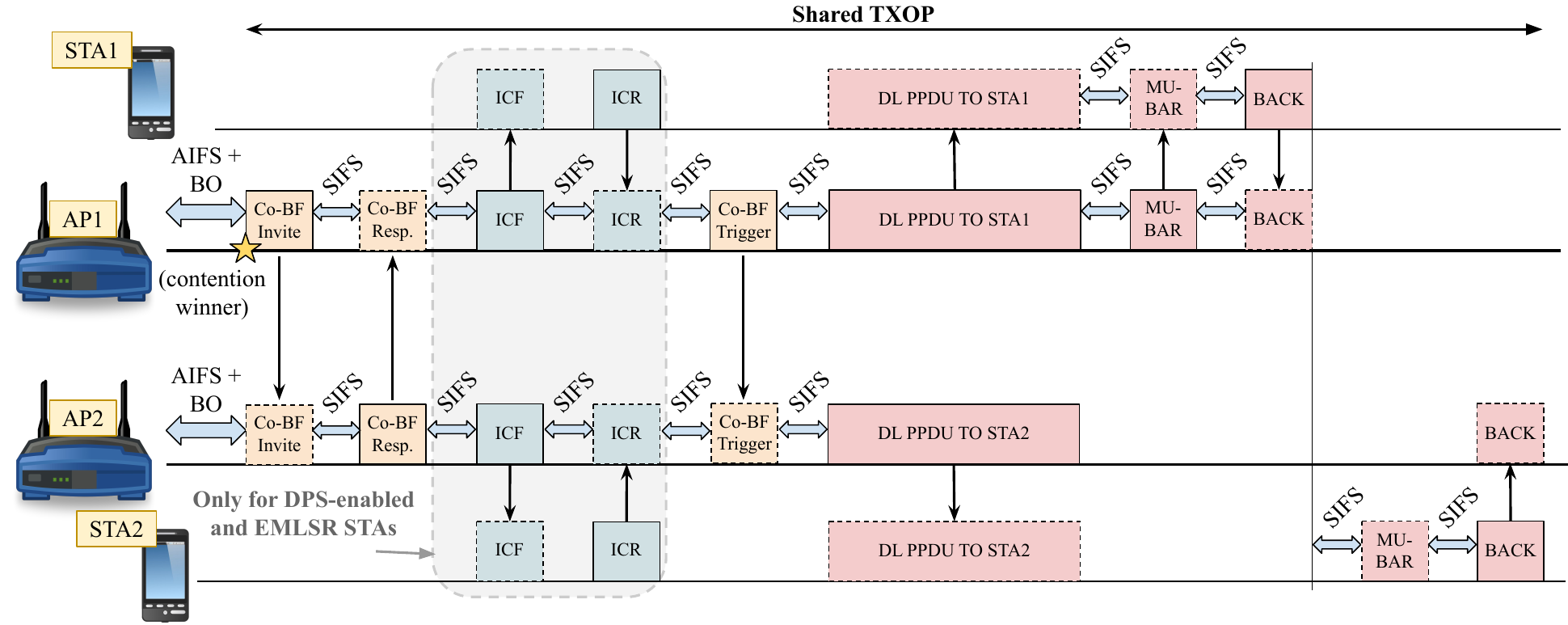}%
    \label{fig:mapc_features_cobf_2}}
    \caption{\gls{cobf} operation within two coordinated \glspl{bss}. (a) Example deployment, (b) \gls{cobf} simultaneous transmission.}
    \label{fig:mapc_features_cobf}
\end{figure*}

\section{IEEE 802.11bn MAPC Features}
\label{sec:mapc_features}

This section comprehensively describes the five candidate features for \gls{mapc} in 11bn, namely \gls{cobf}, \gls{cosr}, \gls{cotdma}, \gls{cortwt}, and \gls{cocr}, showing their specific functionalities and discussing their respective literature. As an important remark, both \gls{cobf} and \gls{cosr} are limited to downlink transmissions only, whereas the rest of the schemes do not have that limitation. Nevertheless, for the sake of illustration, all the examples provided consider only downlink transmissions.


\subsection{Coordinated Beamforming (Co-BF)}

\gls{cobf} is an \gls{mapc} feature that aims to improve spectral efficiency by carefully managing inter-\gls{bss} interference during simultaneous downlink transmissions, potentially leading to higher performance and reliability. For that, it leverages beamforming null steering through antenna arrays, whereby each coordinated \glspl{ap} can minimize the interference measured at each other's recipient \gls{sta} by directing spatial nulls towards them \cite{GerGarLop2017,GarGerGal2018}. In 11bn, the number of simultaneous \gls{cobf} transmissions is limited to two \glspl{ap}, the maximum number of spatial streams per receiving \gls{sta} is restricted to two, and the maximum total number of spatial streams among all recipients is four. In addition, the total number of recipient \glspl{sta} across the two \glspl{ap} in a \gls{cobf} transmission is limited to four.

\textbf{Operation:} The beamforming null steering technique applied in \gls{cobf} is sketched in Fig.~\ref{fig:example_beamforming}, where two \glspl{ap} (\gls{ap}~1 and \gls{ap}~2) employ an antenna pattern during their transmissions to maximize the gain towards their recipient \gls{sta}, but at the same time minimize the interference caused at the \gls{sta} associated with the other coordinated \gls{ap}. For each \gls{ap}, the agreed antenna pattern includes a main lobe and a null, which are steered to the spatial direction of the corresponding targets. For instance, for \gls{ap}~1, the main lobe is steered to \gls{sta}~1 (at a 270$^\circ$ angle) and the null to \gls{sta}~2 (at a 45$^\circ$ angle).\footnote{The desired antenna patterns are in practice achieved with techniques like zero-forcing precoding~\cite{yoo2006optimality}.} In Wi-Fi, the necessary information for generating the antenna patterns is extracted from \gls{csi} matrices, which contain the complex gain and phase shifts from every transmit antenna to every receive antenna at the \gls{sta} over the entire bandwidth. The collection of \gls{csi} requires sounding the channel to learn its characteristics, which involves signaling (e.g., \glspl{ndp} exchanges, transmission of compressed \gls{csi} feedback) between the \gls{ap} and the \glspl{sta}~\cite{cao2024revisiting}. In 11bn, these procedures are extended to allow coordinated \glspl{ap} to collect the necessary cross-\gls{bss} \gls{csi} and to achieve the synchronization needed to perform simultaneous \gls{cobf} transmissions.

\begin{figure*}[ht!]
    \centering    
    \includegraphics[width=0.75\textwidth]{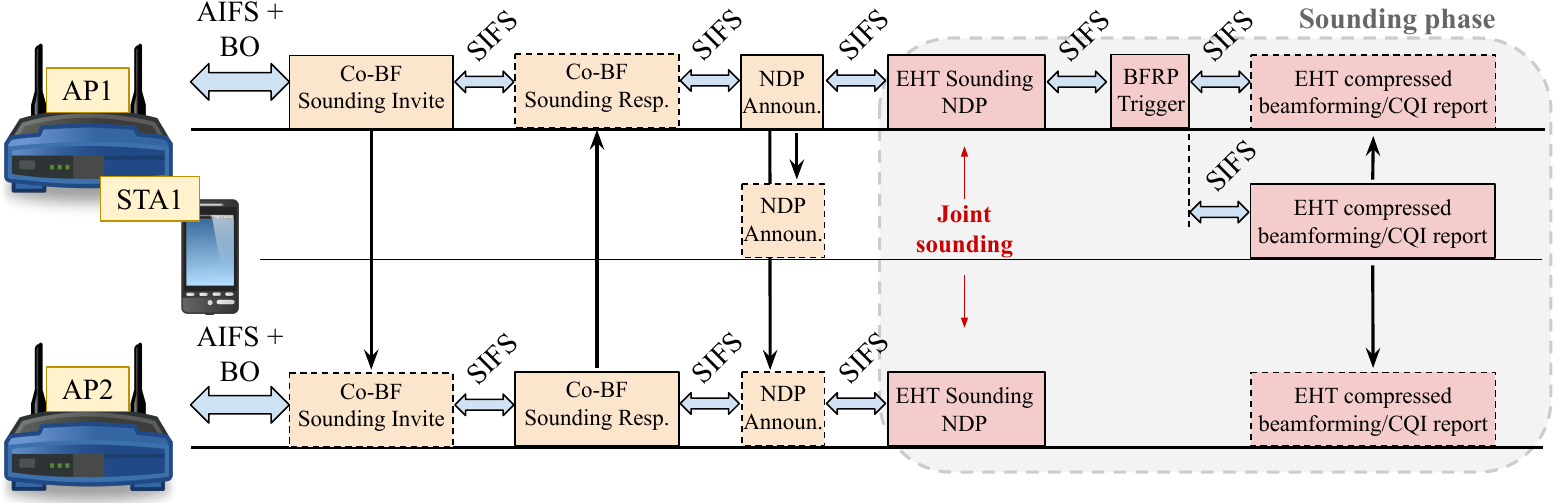}    
    \caption{\gls{cobf} joint sounding procedure initiated by \gls{ap}~1. Note that the counterpart sequence to collect \gls{sta}~2's \gls{csi} (initiated by \gls{ap}~2 and needed to compute the joint beamforming/null-steering precoder) is omitted for the sake of space, but the procedure is the same as for \gls{sta}~1.
    }
    \label{fig:cobf_sounding}
\end{figure*}

\textbf{Setup:} For the discovery of \gls{cobf} capabilities, a \gls{cobf} subelement is included in the \gls{mapc} Parameter Set (see Fig.~\ref{fig:mapc_parameter_set} from Appendix~\ref{sec:appendix_a}) to inform about: \emph{i)} the maximum number of supported sounding \gls{obss} reports by the \gls{ap}, \textit{ii)} the number of joint sounding reports that the \gls{ap} can store for the \gls{cobf} pair, and \emph{iii)} the \gls{ap}'s support for using the combination of \texttt{2$\times$LTF} (the sequence in \gls{ltf} is duplicated to provide better channel estimation accuracy) and \texttt{0.8~$\mu$s GI} (a shorter \gls{gi} is used compared to standard 1.6~$\mu$s or 3.2~$\mu$s values) during \gls{cobf}. Then, the \gls{cobf} agreement negotiation is performed following the procedure described in Section~\ref{sec:mapc_negotiation}.

\textbf{Procedure:} With a \gls{cobf} agreement in place, the realization of simultaneous \gls{cobf} transmission is enabled by a coordinated sounding operation and the exchange of \gls{csi} data between the \glspl{ap}. \gls{cobf} transmissions and channel sounding are decoupled from each other.\footnote{The frequency at which an \gls{ap} performs channel sounding is implementation-specific and depends on how fast and how much the channel varies in a given environment (e.g., triggered by \gls{snr} degradation).} This sounding procedure reuses and extends the one defined in 802.11be (\gls{eht}) and is performed as follows (illustrated in Fig.~\ref{fig:cobf_sounding}):
\begin{enumerate}
    \item An initiating (coordinating) \gls{ap} sends a \texttt{\gls{cobf} Sounding Invite}, indicating the sounding capabilities (e.g., number of supported spatial streams) of the involved \glspl{sta}.
    \item After a \gls{sifs} time, the responding (coordinated) \gls{ap} sends a \texttt{\gls{cobf} Sounding Response} to indicate the acceptance or rejection of the request.
    \item If accepted, the requesting \gls{ap} sends a \texttt{\gls{uhr} \gls{ndpa}}. This frame contains information to indicate the \glspl{ap} expected to transmit sounding sequences and the set of \glspl{sta} that are expected to provide channel feedback.
    \item From this point, either a sequential or a joint \gls{ndp} sounding phase takes place. For the joint case (the one illustrated in Fig.~\ref{fig:cobf_sounding}), both the initiating and responding \glspl{ap} transmit an \textit{\gls{eht} Sounding \gls{ndp}} to the beamformee (i.e., the \gls{sta} that is going to compute the \gls{csi} matrix). While two \glspl{ndp} are transmitted simultaneously on the same frequency channel, independent per-\gls{ap} channel estimation can be achieved in the joint procedure. In particular,  each \gls{ap} uses orthogonal and non-overlapping columns of the \gls{ltf} training matrix, allowing the beamformer to separate each \gls{ap}'s contribution from the superposed \gls{ndp}.    
    \item After the \gls{ndp} sequence, the initiating \gls{ap} sends a \texttt{\gls{bfrp}} \texttt{Trigger} to gather the \gls{csi} information (in the form of a compressed beamforming feedback matrix) from the beamformee.
    \item Finally, the beamformee replies with an \texttt{\gls{eht}} \texttt{\gls{tb}} \texttt{\gls{ppdu}} containing the compressed beamforming report, which is acquired by both the initiating and responding \glspl{ap}. 
\end{enumerate}

Once \gls{csi} is available, the realization of simultaneous transmissions is done through the following steps (shown in Fig.~\ref{fig:example_beamforming}):
\begin{enumerate}
    \item An \gls{ap} that wins a \gls{txop} takes the role of \textit{coordinating \gls{ap}} and can invite another \textit{coordinated} \gls{ap} to perform a simultaneous \gls{cobf} transmission.
    \item The coordinating \gls{ap} sends a \texttt{\gls{cobf} Invite}, which is included in a \texttt{\gls{bsrp} \gls{ntb}} frame and includes the \gls{mac} addresses of both coordinating and coordinated \glspl{ap} and relevant parameters (e.g., min./max. number of data \gls{ofdm} symbols, bandwidth used, \gls{gi} and \gls{ltf} size, number of spatial streams).
    \item After a \gls{sifs} time, the coordinated \gls{ap} sends \texttt{\gls{cobf} Response}, which is included in a \texttt{multi-\gls{sta} \gls{back}}, to accept or decline the invitation. If accepted, it confirms the parameters indicated by the coordinating \gls{ap} or suggests new ones. If rejected, a reason code is provided.
    \item Only in some circumstances---i.e., when \glspl{sta} use \gls{dps}, for \gls{emlsr} transmissions, or for \glspl{sta} using \gls{duo}---an \gls{icf}-\gls{icr} frame exchange between each \gls{ap} and its recipient \gls{sta} is required before the \gls{cobf} trigger that precedes the simultaneous \gls{cobf} transmission. The \gls{icf}-\gls{icr} exchange allows waking up the \gls{sta} from a power-saving mode in \gls{dps}, and keeping the \gls{sta} on a given link (to avoid premature link switching) for \gls{emlsr}. 
    \item After a \gls{sifs} time, the coordinating \gls{ap} sends a trigger frame to confirm the final parameters for the simultaneous transmission (e.g., bandwidth used, \gls{mcs} of each \gls{sta}, \gls{txop} duration). To ensure time and frequency synchronization between the \glspl{ap}, the trigger frame includes the final set of parameters to construct a common preamble for the subsequent data \glspl{ppdu} (e.g., \gls{lsig} length, \gls{gi}, \gls{ltf} type, number of \gls{uhr}-\gls{ltf} symbols).    
    \item The simultaneous \gls{ppdu} transmission(s) takes place, with each \gls{ap} potentially serving multiple \glspl{sta} via \gls{mu}-\gls{mimo}.
    \item Finally, a sequential \gls{ack} mechanism takes place, where the involved \glspl{ap} send a \texttt{\gls{mu}-\gls{bar} Trigger frame} (including the identifier of the recipient \glspl{sta} in the \texttt{User Info} field) to explicitly solicit a \gls{back} (included in an \texttt{\gls{uhr} Trigger-Based \gls{ppdu}}) from the recipient \glspl{sta}. This operation is done sequentially, being the coordinating \gls{ap} first and the coordinated \gls{ap} second (after $2\times$SIFS plus the coordinating \gls{ap}'s \gls{ack} duration). In case \textit{HETP ACK} is used, which is the mechanism for acknowledging \gls{he} \gls{tb} \glspl{ppdu}, no \gls{mu}-\gls{bar} is needed, and the \gls{back} is directly transmitted.
\end{enumerate}

\textbf{Literature on \gls{cobf}:} The fact that \gls{cobf} was previously a candidate feature for Wi-Fi~7 led to early evaluation works like~\cite{garcia2021ieee}, where it was shown to provide significant gains in terms of worst-case latency. Now, with Wi-Fi 8 and \gls{uhr} approaching, \gls{cobf} has become an important feature to precisely meet those reliability goals~\cite{galati2024will}. Due to its inherent complexity and associated overheads, \gls{cobf} has also been extensively investigated within the 802.11 standardization process in order to validate its efficacy~\cite{garcia2020performance, lopez2022performance, li2023performance}. A critical aspect behind \gls{cobf} lies in the trade-off it poses between its required overheads and the performance gains it may bring along. Several works have studied this trade-off~\cite{you2025improving, aboushehada2026autoencoder}, supporting the idea that \gls{cobf} is not unconditionally beneficial, especially at low-to-moderate loads or in regimes where the channel varies very fast (e.g., due to \gls{sta} mobility). Accordingly, the success of \gls{cobf} might be tied to a proper management layer that decides under which conditions it must be applied.  

\subsection{Coordinated Spatial Reuse (Co-SR)}

\gls{cosr} is an \gls{mapc} feature that, similar to \gls{cobf}, aims to enable simultaneous downlink transmissions by different \glspl{ap}. To do it, a simpler approach based on power control is adopted. By targeting improved spectrum utilization, \gls{cosr} also aims for increased network throughput and reliability, but at a lower cost than \gls{cobf} (e.g., \gls{csi} acquisition is not required to implement \gls{cosr}).

\begin{figure}[ht]
\centering
\includegraphics[width=0.8\columnwidth]{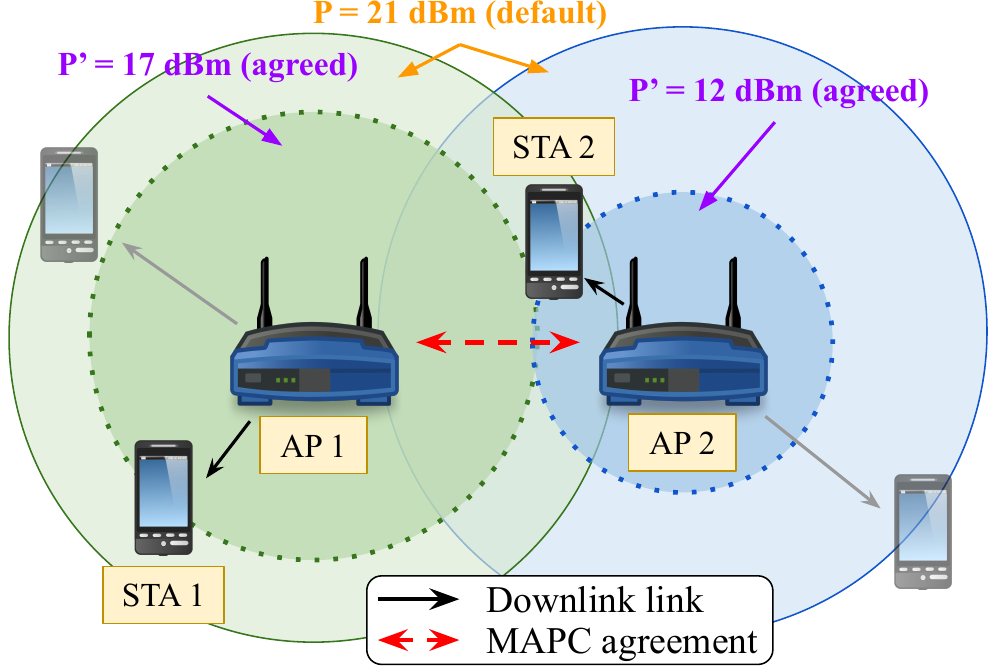}
\caption{\gls{cosr} operation within two coordinated \glspl{bss}.}
\label{fig:cosr_1}
\end{figure}

\textbf{Operation:} Fig.~\ref{fig:cosr_1} depicts an example deployment in which two coordinated \glspl{bss} apply \gls{cosr}. To perform a simultaneous \gls{cosr} transmission, both \glspl{ap} agree on the transmit power to be used (e.g., $P_1=17$~dBm and $P_2=12$~dBm for \gls{ap}~1 and \gls{ap}~2, respectively), so that the interference at the recipient \glspl{sta} (\gls{sta}~1 and \gls{sta}~2) does not degrade the respective reception qualities. This differs from preceding decentralized approaches like 11ax \gls{obsspd}-based \gls{sr}, where the transmit power is tuned as a result of relaxing the channel access through the \gls{obsspd} threshold. In a \gls{cosr} transmission, instead, the coordinating \gls{ap} decides its own transmit power and the maximum allowed transmit power to be used by the coordinated \gls{ap} in a simultaneous transmission. The standard does not define any specific algorithm to compute the transmit power of both the coordinating and the coordinated \glspl{ap}, and leaves its implementation open. Nevertheless, \gls{rssi} measurements from \gls{obss} neighbors taken by the \gls{sta} can be useful to derive such a transmit power limit (e.g., in a way that the expected \gls{sinr} does not fall below a given threshold). It is worth noting that, in practice, deriving the transmit power in \gls{cosr} requires knowledge of the interference path between the coordinated \glspl{ap} and their \glspl{sta}. The exchange of such information is not included within the \gls{mapc} framework and must be obtained through other mechanisms like 802.11k neighbor reports.

\textbf{Setup:} The \gls{cosr} agreement is managed via the \gls{mapc} negotiation procedures described in Section~\ref{sec:mapc_negotiation}. For \gls{cosr} in particular, the \gls{cosr} profile within the MAPC element is used throughout the negotiation. Such a profile includes the minimum transmit power the coordinated \gls{ap} is willing to use, ensuring that the forthcoming simultaneous transmission is worth it.

\textbf{Procedure:} \gls{cosr} follows a very similar procedure to \gls{cobf} to orchestrate simultaneous transmissions, but fit for purpose, as summarized next:
\begin{enumerate}
    \item The \gls{txop} winner takes the role of \textit{coordinating \gls{ap}}.
    \item The coordinating \gls{ap} sends a \texttt{\gls{cosr} Invite}, which includes relevant parameters like the min./max. number of data \gls{ofdm} symbols, need or not for \gls{icf}-\gls{icr} frame exchange, or the duration of such an exchange. 
    \item The coordinated \gls{ap} sends \texttt{\gls{cosr} Response}.
    \item When needed (any scheduled \gls{sta} uses \gls{dps} or \gls{emlsr}), an \gls{icf}-\gls{icr} frame exchange is performed.
    \item Before the simultaneous transmission, the coordinating \gls{ap} sends a \gls{cosr} trigger and confirms the agreed parameters, e.g., duration of the data \gls{ppdu} for both \glspl{ap}, the transmit power of the coordinating \gls{ap}, the transmit power limit of the coordinated \gls{ap} in dBm/20 MHz.  
    \item The simultaneous \gls{ppdu} transmission takes place. Two modes are defined: \emph{Mode 1} (allows either \gls{uhr} and \gls{eht} \glspl{ap} transmissions) and \emph{Mode 2} (only for \gls{uhr} \glspl{ppdu} transmissions).
    \item The transmission of \glspl{back} is triggered by the coordinating \gls{ap}.
\end{enumerate}

\textbf{Literature on \gls{cosr}:} \gls{cosr} evolves the concept of \gls{sr}, introduced in 802.11ax (11ax), to enhance the spectral efficiency of a given geographical area through simultaneous transmissions~\cite{wilhelmi2021spatial}, by leveraging the coordination between neighboring \glspl{ap}. The coordination enabled by \gls{mapc} is instrumental to addressing the issues associated with 11ax \gls{sr}, where the adaptation of the transmission power is carried out independently by different \glspl{ap}, without knowledge of the real impact of such decisions. However, given the gap posed by 11bn---which does not define how to compute the best transmit power, nor standardizes mechanisms to exchange relevant information such as \gls{rssi}---some research works have proposed various mechanisms and evaluations to show the potential performance gains of \gls{cosr} ~\cite{talukder2023enhanced, nunez2022txop, imputato2024beyond, wilhelmi2023throughput, nunez2024spatial, jung2025coordinated}. When it comes to approaches to computing the power, the dominant approach is based on deriving the power limits from \gls{rssi} measurements of \gls{obss} transmissions collected by \glspl{sta}. However, the \gls{rssi} represents an average measurement rather than an instantaneous value, so it might fail to capture real-time phenomena such as small-scale fading or mobility. In addition, keeping these measurements fresh across \glspl{bss} entails signaling overheads, which were studied in~\cite{nunez2024spatial}. Finally, another inherent challenge of \gls{cosr} is the formation of \gls{cosr} groups to carry out simultaneous transmissions, for which a complex set of aspects must be considered, including the geometric characteristics of deployment, traffic patterns, or prioritization requirements~\cite{nunez2023multi}. Together, these challenges motivate the use of learning-based mechanisms to decide when and with whom to apply \gls{cosr}. The authors in \cite{wilhelmi2024coordinated} propose coordinated multi-armed bandits that learn the best spatial reuse policies by combining power and packet-detection threshold configurations across multiple coexisting BSSs, showing significant gains over uncoordinated baselines. In \cite{wojnar2024ieee}, a hierarchical multi-armed bandits for \gls{cosr} group selection is proposed, showing that data-driven solutions offer rapid convergence, adaptability to topology changes, and sustained performance. In \cite{nunez2025deep}, deep reinforcement learning is used to perform \gls{cosr} scheduling, minimizing the worst-case latency and better generalization properties than established scheduling heuristics.

\subsection{Coordinated Time-Division Multiple Access (Co-TDMA)}

\gls{cotdma} is an inter-\gls{ap} scheduling feature that aims to provide collision-free channel access by distributing a \gls{txop} among the different participating \glspl{ap}. The main objective behind \gls{cotdma} is to mitigate some of the limitations of the randomized channel access from the \gls{dcf} (e.g., excessive contention).

\begin{figure*}[ht!]
    \centering
    \subfloat[]{\includegraphics[width=0.25\textwidth]{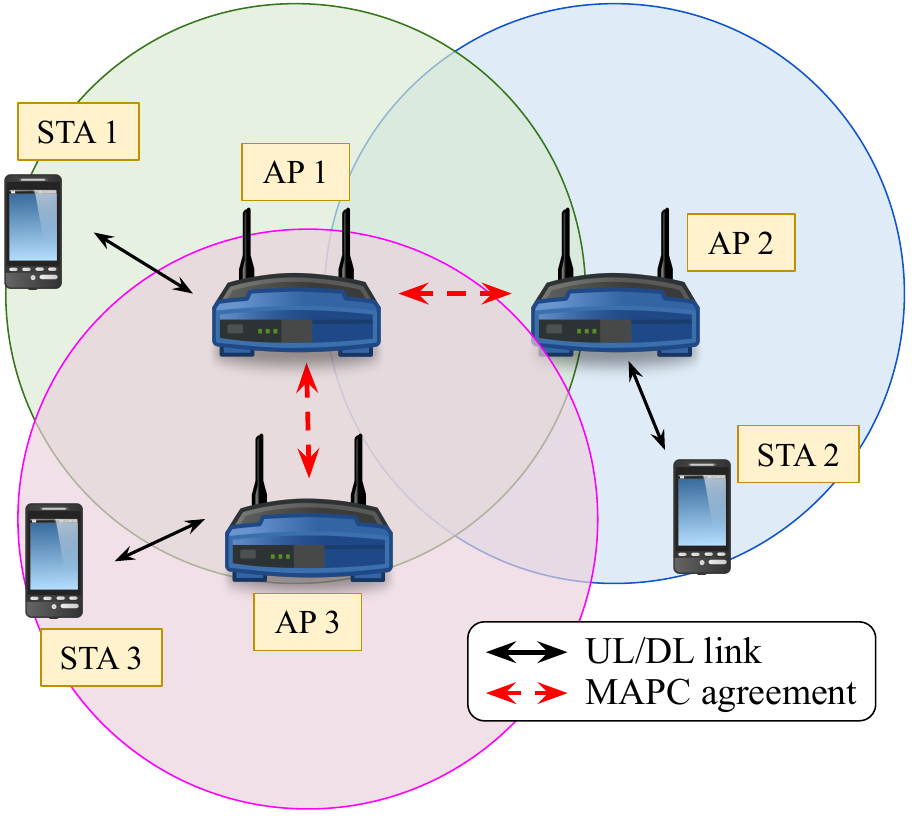}%
    \label{fig:example_cotdma}}
    \hfil
    \subfloat[]{\includegraphics[width=0.75\textwidth]{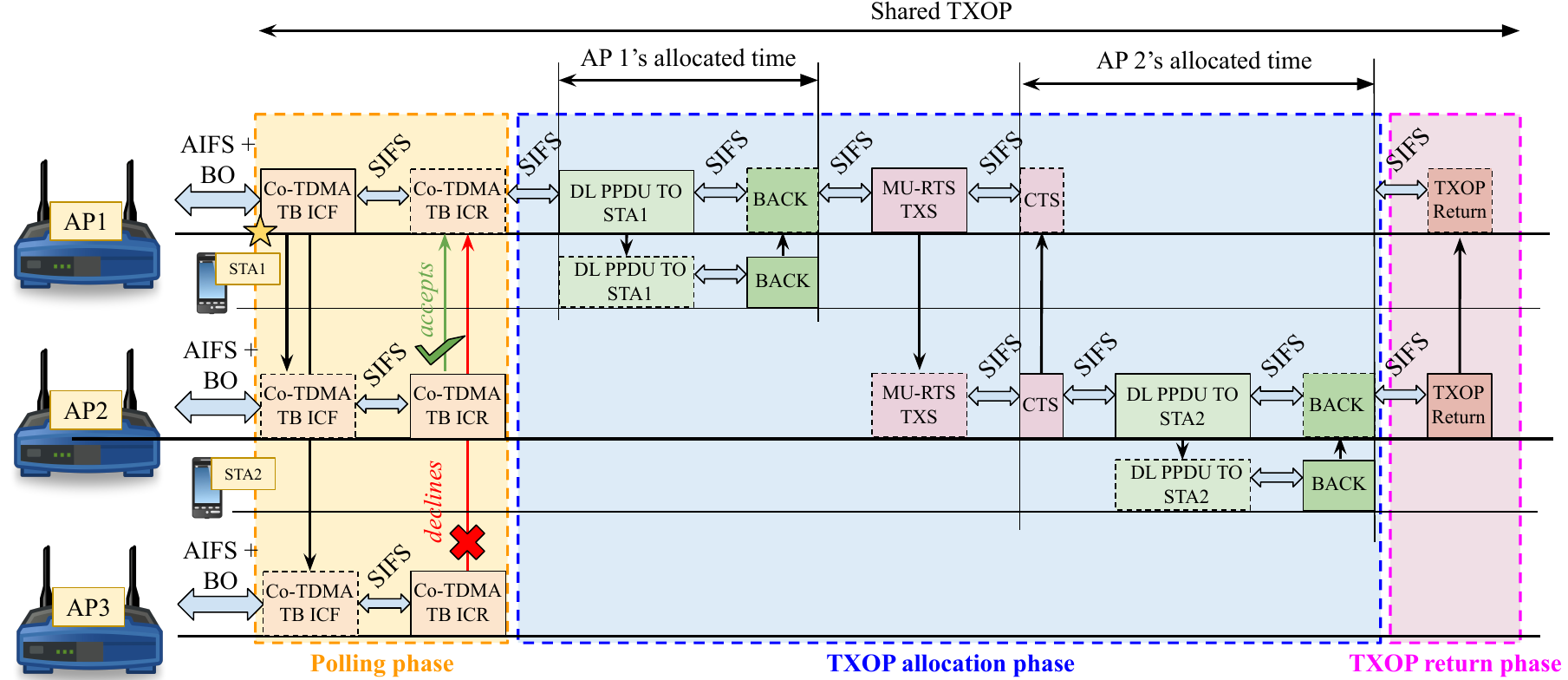}%
    \label{fig:mapc_features_cotdma_2}}
    \caption{\gls{cotdma} operation within three coordinated \glspl{bss}. (a) Example deployment, (b) Scheduled \gls{cotdma} transmissions.}
    \label{fig:mapc_features_cotdma}
\end{figure*}

\textbf{Operation:} In \gls{cotdma}, a coordinating \gls{ap} can allocate portions of a \gls{txop} to one or more coordinated \glspl{ap} (sharing the same primary 20~MHz channel) to hold collision-free transmissions to their associated \glspl{sta}. On top of that, \gls{cotdma} needs to be compatible with legacy \gls{dcf}. For that, an \gls{ap} running \gls{cotdma} must reserve a portion of the time for contention-based channel access. In \gls{cotdma}, \gls{edca} traffic prioritization is enforced by exclusively allocating time resources to traffic of equal or higher priority than the primary \gls{ac} of the acquired \gls{txop}. For example, a \gls{txop} won under AC\_BE can be used for AC\_BE, AC\_VI, or AC\_VO, but not for AC\_BK. This prevents low-priority traffic from gaining preferential channel access under the cover of a high-priority \gls{txop}, preserving \gls{qos} differentiation.

\textbf{Setup:} As for the rest of the features, 
the negotiation and setup of \gls{cotdma} agreements work as described in Section~\ref{sec:mapc_negotiation}. 
During the setup of a \gls{cotdma} agreement, the involved \glspl{ap} need to ensure there is a \gls{bss} bandwidth overlap. 
For this, the \gls{cotdma} profile is exchanged, which includes a bandwidth control field to inform about both the channel identity (channel center frequency index) and the bandwidth configuration (encoding 20/40/80/160/320 MHz) of the channels to be used (see Fig.~\ref{fig:cotdma_profile} from Appendix~\ref{sec:appendix_a}). In \gls{cotdma}, despite the limit to establishing agreements between only two \glspl{ap} in 11bn, multiple coordinated \glspl{ap} (each with a separate agreement) can be scheduled during \gls{cotdma} \glspl{txop}.

\textbf{Procedure:} \gls{cotdma} defines three phases, referred to as \textit{polling} (to check interest from coordinated \glspl{ap}), \textit{\gls{txop} allocation} (where scheduled transmissions are held), and \textit{\gls{txop} return} (whereby coordinated \glspl{ap} can return unspent allocated time to the coordinating \gls{ap}). Overall, the procedure is as follows (see Fig.~\ref{fig:mapc_features_cotdma}):
\begin{enumerate}
    \item A coordinating \gls{ap} wins a \gls{txop} and decides to share a portion of it using \gls{cotdma}. 
    \item The coordinating \gls{ap} sends an \gls{icf} frame (included in a \texttt{\gls{bsrp} Trigger} frame) to poll one or more coordinated \glspl{ap}. As part of \gls{cotdma}, the coordinating \gls{ap} can poll non-\gls{ht} \gls{ppdu} by sending an \gls{ntb} \gls{icf}. The coordinating \gls{ap} has the authority over the allocation of time resources, and the proposed slots can be accepted or rejected by the coordinated \glspl{ap}.
    \item The polled \glspl{ap} reply either positively or negatively to receiving a time allocation within the upcoming \gls{txop} by sending a \gls{cotdma} \gls{icr} (included in a \texttt{multi-\gls{sta} \gls{back}}).
    \item Once the allocation of resources is clear, the coordinating \gls{ap} triggers coordinated transmissions by transmitting \texttt{\gls{mu}-\gls{rts} \gls{txs} Trigger} frames, which include the identifier of the \gls{ap} to transmit next, the allocated time, and channel(s).
    \item When triggered, the coordinated \gls{ap} first confirms the \gls{txop} by sending a \texttt{\gls{cts}} and then uses its allocated time to transmit one or more \glspl{ppdu}. 
    \item Finally, the coordinated \gls{ap} can finish its slot earlier and release the resources by sending an \texttt{\gls{mapc} \gls{txop} Return} frame (for which no response is expected), allowing the coordinating \gls{ap} to schedule more transmissions from other \glspl{ap}.    
\end{enumerate}

\textbf{Literature on \gls{cotdma}:} The performance of \gls{cotdma} has been significantly studied in~\cite{val2025wi, lee2025performance, cai2020coordinated}, showing its benefits in terms of reliability and latency in high-density deployments. The allocation of \gls{cotdma} resources in centralized controller architectures has also been studied in~\cite{zhang2025modeling}, where an optimal transportation theory approach was taken. Finally, to address the moderated gains provided by \gls{cotdma} in some scenarios (e.g., in small/mid deployments where \glspl{sta} are close to their \gls{ap}~\cite{nunez2022txop}), the works in~\cite{nunez2025improving, imputato2024beyond} also evaluated \gls{cotdma} in conjunction with \gls{cosr} (highlighting mutual benefits of combining both schemes) and by allowing the formation of groups with more than two \glspl{ap}, showing the future potential of \gls{cotdma}. 

Despite its promise for latency reduction in dense deployments, \gls{cotdma} also faces several practical challenges. First, its associated overhead (\gls{icf}-\gls{icr} exchange) consumes airtime before any data is transmitted, which is detrimental in low-to-moderate load regimes. Second, although \glspl{ap} exchange information about \gls{scs} during negotiation (so that they know \gls{txop} requirements of each other in advance), the allocation of time resources entails a complex decision-making problem with partial information (the coordinating \gls{ap} does not have knowledge of the coordinated \gls{ap}'s queue depth or traffic) that involves aspects such as transmission quality and prioritization. Later in this paper (Section~\ref{sec:simulations}), we assume an equal split of time resources, which is shown to be suboptimal. Third, \gls{ac} prioritization constraints the utilization of slots by lower-priority traffic than the primary \gls{ac} of the \gls{txop}, creating a strong rigidity in mixed-traffic scenarios~\cite{lee2025performance}. Finally, \gls{cotdma} can penalize non-coordinated \glspl{bss} by increasing their effective backoff times, as also confirmed by our simulation results. Note that the 11bn specification includes fairness rules for \gls{cotdma} (\gls{txop} sharing cap and contention-based access reservation) to avoid penalizing other \glspl{bss}. Together, these challenges motivate the introduction of intelligent scheduling mechanisms that decide when to activate \gls{cotdma} and how to split time resources~\cite{zhang2025modeling, nunez2025deep}.

\subsection{Coordinated Restricted Target Wake Time (Co-RTWT)}

\gls{cortwt} extends the \gls{rtwt} feature introduced in 11be to allow multiple \glspl{ap} to coordinate their \gls{rtwt} schedules, thus adding additional inter-\gls{bss} protection to such operations. With this, \gls{cortwt} aims to reduce \gls{mpdu} losses and provide support to time-sensitive traffic.

\textbf{Operation:} In \gls{rtwt}, both the \gls{ap} and its associated \glspl{sta} can negotiate \gls{rtwt} agreements whereby \gls{sp} can be reserved for one or more \glspl{sta}. An \gls{rtwt} agreement specifies the start time, duration, and periodicity of an \gls{sp}. Consequently, no other transmissions except the scheduled ones can occur during \glspl{sp}. In \gls{cortwt}, this whole operation is extended to ensure the protection from other \glspl{bss}, as detailed next.

\begin{figure*}[t!]
    \centering
    \subfloat[]{\includegraphics[width=0.3\textwidth]{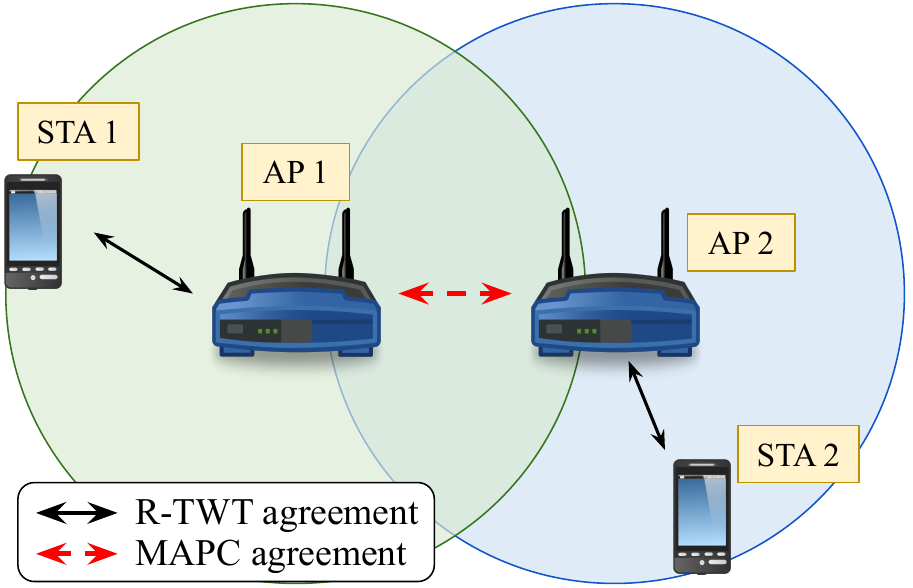}%
    \label{fig:mapc_features_cortwt_1}}
    \hfil
    \subfloat[]{\includegraphics[width=0.7\textwidth]{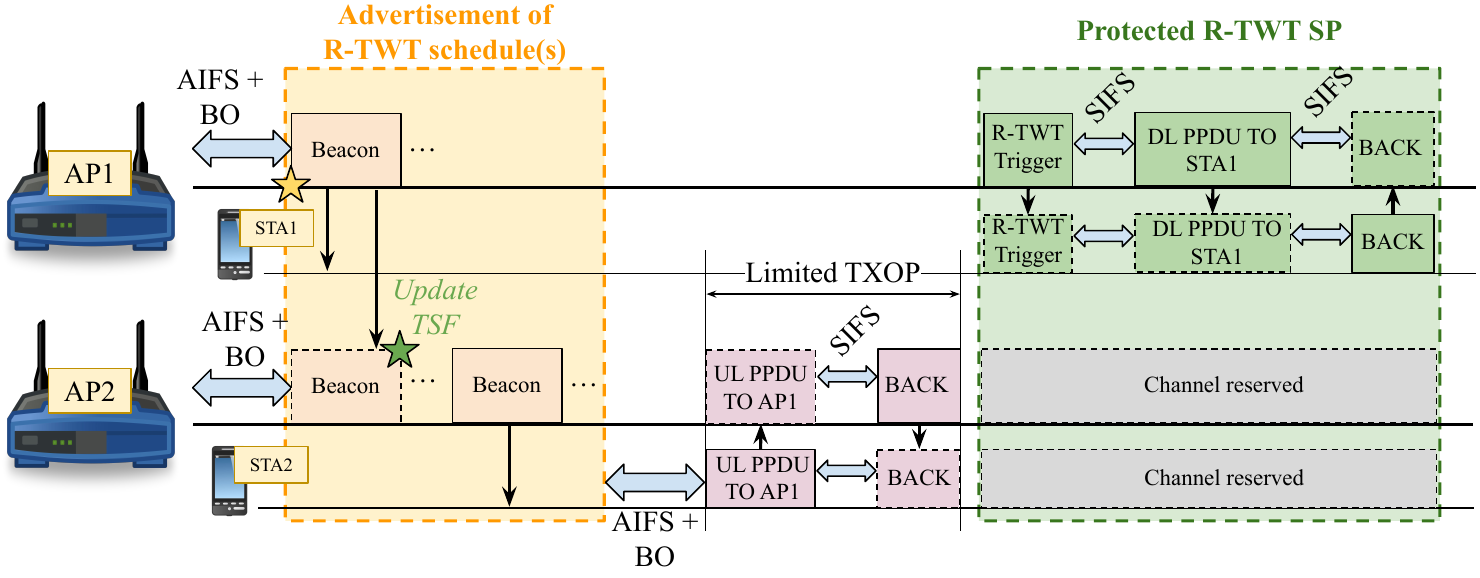}%
    \label{fig:mapc_features_cortwt_2}}
    \caption{\gls{cortwt} operation within two coordinated \glspl{bss}. (a) Example deployment, (b) Extended \gls{rtwt} \gls{sp} protection.}
    \label{fig:mapc_features_cortwt}
\end{figure*}

\textbf{Setup:} The negotiation and setup of \gls{cortwt} agreements are based on the exchanged information using the \gls{cortwt} profile of the \texttt{\gls{mapc} Element}, which provides detailed information about the individual \gls{rtwt} schedules to be protected (see Fig.~\ref{fig:cortwt_profile} from Appendix~\ref{sec:appendix_a}). \gls{cortwt} agreements are uniquely identified by a broadcast \gls{twt} ID and the \gls{mac} address of the requesting \gls{ap}.

\textbf{Procedure:} Unlike the previously overviewed schemes, \gls{cortwt} does not operate on a per-\gls{txop} basis. Instead, it focuses on providing inter-\gls{bss} protection for the \gls{rtwt} schedules. For that reason, we differentiate between \textit{\gls{cortwt} requesting \glspl{ap}} (which request \gls{rtwt} schedule protection) and \textit{\gls{cortwt} responding \glspl{ap}} (which agree on such a protection). In particular, this is done as follows (see Fig.~\ref{fig:mapc_features_cortwt}):
\begin{enumerate}
    \item The coordinating \gls{ap} advertises its active \gls{rtwt} schedules by including them into the \textit{Restricted TWT Parameter Set} fields of the \gls{twt} element, which are transmitted in Beacon frames.
    \item The coordinated \gls{ap} converts the requested \glspl{twt} to its own local \gls{tsf}, ensuring that all time parameters are accurately aligned with the local BSS timer. This allows announcing (via Beacon frames) the updated schedules to the \glspl{sta} of its \gls{bss}, which will apply the protection.
    \item Before initiating any transmission, the coordinated \gls{ap} must ensure that the transmission will be performed before the next protected \gls{cortwt} \gls{sp} starts. If not possible, then it must defer its backoff.
    \item When holding a \gls{txop}, the coordinated \gls{ap} must ensure that its intra-\gls{bss} transmissions finalize before the start of a protected \gls{cortwt} \gls{sp}.
    \item During a scheduled \gls{rtwt} \gls{sp}, the coordinating \gls{ap} (optionally) sends a Trigger frame to explicitly indicate the data exchange to the corresponding \gls{sta}(s) and add further channel protection.
\end{enumerate}

\textbf{Literature on \gls{cortwt}:} The performance of \gls{cortwt} has been barely studied in the literature~\cite{peng2024access}. However, the potential gains of \gls{rtwt} have already been demonstrated for single \gls{bss} deployments, showing the paradigmatic shift of the feature in terms of reliability and support for low-latency traffic compared to its predecessor, \gls{twt}, which instead focused on energy saving~\cite{nurchis2019target, mozaffariahrar2025r, haxhibeqiri2024coordinated, belogaev2024dedicated}. In addition, the need for coordination in \gls{rtwt} had already been discussed during the Wi-Fi 7 standardization period~\cite{bankov2024performance}, advocating for cases where \glspl{ap} serve latency-sensitive and deterministic traffic (e.g., industrial \gls{iot}, real-time control, \gls{xr}). 

In terms of challenges, the effectiveness of \gls{cortwt} depends on the curated design of \gls{rtwt} schedules to match the actual traffic demand and requirements. For instance, wasting protected airtime by overprovisioning \glspl{sp} for a given traffic source or class negatively affects the overall performance, which has been shown to be particularly exacerbated under bursty traffic~\cite{mozaffariahrar2025r}. Moreover, as shown in \cite{ramanna2025tcp}, the interplay between \gls{rtwt} and higher-layer communication protocols like \gls{tcp} may lead to counterproductive dynamics (e.g., \gls{tcp} \gls{ack} timeouts can halt and decrease performance due to \gls{rtwt}) if not properly addressed. Translating these intra-\gls{bss} challenges into \gls{cortwt} leads to further complexity, which motivates the usage of adaptive mechanisms that activate and configure \gls{cortwt} based on observed activity, interference levels, and traffic characteristics.

\subsection{Coordinated Channel Recommendation (Co-CR)}

\gls{cocr} goes hand in hand with another Wi-Fi 8 feature, \gls{txspg}, whereby a portion of a \gls{txop} can be allocated to \gls{p2p} traffic. This is particularly useful for networks where devices need to communicate locally (e.g., a \gls{vr} device communicating with a game console) without going through the \gls{ap}. In particular, \gls{cocr} is centered on optimizing those \gls{p2p} communications, and it does so by making \glspl{ap} agree on the channels used for that purpose, thus cleanly separating the transmissions from different \gls{p2p} groups. In this case, the goal is to make \gls{p2p} channel choices cooperatively, granting protection, predictability, and stability to \gls{p2p} connections. Unlike for the rest of the schemes, \gls{cocr} has no dedicated literature to date, hence we next focus on its operation, setup, and procedure only.

\textbf{Operation:} \gls{cocr} simply enhances the \gls{txspg} operation by allocating \gls{p2p} among different \glspl{bss} optimally. In \gls{txspg}, the \gls{ap} allocates a portion of \gls{txop} for \gls{p2p} transmissions. The awareness of \gls{p2p} traffic needs is obtained thanks to \gls{scs}, a procedure whereby \glspl{sta} can request \gls{qos} for specific data streams. In this context, a \gls{txspg} requesting \gls{sta} can send an \gls{scs} request to the \gls{ap} to indicate the need for transmitting \gls{p2p} traffic. This request includes an \texttt{\gls{scs} Descriptor} element, which identifies the flow, the minimum and maximum required service interval, a delay bound, and a start time. The \gls{ap} uses the \gls{scs} descriptor to allocate the necessary resources, so that the \gls{qos} requirements are met. In subsequent \glspl{txop}, the \gls{ap} can schedule \gls{p2p} transmissions by sending an \texttt{\gls{mu}-\gls{rts} \gls{txs} Trigger} frame.

\textbf{Setup:} Two \glspl{ap} agree on the \gls{cocr} configuration by first exchanging their \gls{cocr} profile (see Fig.~\ref{fig:cocr_profile} from Appendix~\ref{sec:appendix_a}), which the requesting \gls{ap} uses to first provide a list of recommended channels. Optionally, recommendation periods (i.e., time windows when the \gls{p2p} channel is valid) and recommended timeout (i.e., the time in which the channel recommendation has to expire) are provided in time units.

\textbf{Procedure:} Once a \gls{cocr} agreement is in place, the involved \glspl{ap} simply advertise the channels for \gls{p2p} communication through \texttt{Beacon}, \texttt{Probe Response}, and \texttt{(Re)Association Response} frames.

\subsection{On the Interplay of MAPC with Existing Features}

As any new feature in 802.11, \gls{mapc} needs to be carefully designed so that it integrates well with existing features. While some functionalities are independent (e.g., \gls{mcs} design, channelization), others require a thoughtful design. Next, we summarize the main considerations made for \gls{mapc} with respect to backbone 802.11 functionalities:
\begin{itemize}
    \item \textbf{\gls{edca}:} The coordination schemes (e.g., \gls{cobf}, \gls{cosr}, \gls{cotdma}) work atop \gls{edca}, thus no changes are made to the channel access. When an \gls{ap} gains a \gls{txop}, regardless of the \gls{ac}, it becomes a coordinating \gls{ap}, as seen earlier in this section. Nevertheless, when it comes to sharing the acquired \gls{txop}, \gls{ac} awareness is considered, although only for \gls{cotdma}. In particular, the coordinating \gls{ap} signals a \emph{primary \gls{ac}} of the \gls{txop} to inform the involved \glspl{ap} in a \gls{cotdma} transmission about the minimum \gls{ac} priority allowed. Additionally, the total time allocated by the coordinating \gls{ap} to other \glspl{ap} must not exceed the minimum between the primary \gls{ac} \gls{txop} limit and the AC\_VI \gls{txop} limit configured by the coordinating \gls{ap} (notice that if any of those limits is zero, the \gls{txop} shall not be shared).
    \item \textbf{\gls{mlo}:} Despite \gls{mlo} being one of the most important features of 11bn's predecessor (11be), its interaction with \gls{mapc} is minimal. \gls{mapc} operations from \glspl{ap} within the same \gls{ap} \gls{mld} are independent and do not affect each other. In this regard, despite the potential for cross-link coordination, each \gls{ap} is individually responsible for establishing and maintaining its own \gls{mapc} agreements. The only coupling between \gls{mlo} and \gls{mapc} is a consideration in \gls{cobf} and \gls{cosr}, for which coordinated \glspl{ap} need to ensure that \gls{emlsr} \glspl{sta} do not switch links while being part of a transmission. As shown earlier in Fig.~\ref{fig:mapc_features_cobf}, this behavior is prevented through the \gls{icf}-\gls{icr} exchange, which can serve to prevent any transmissions to other links of the target \gls{emlsr} \gls{sta}.
    \item \textbf{Power saving:} The interplay between \gls{mapc} and power saving modes like \gls{dps} is critical, hence it has been intrinsically considered for the design of \gls{mapc} transmission exchanges. In \gls{cobf}, \gls{cosr}, and \gls{cotdma}, the \gls{icf}-\gls{icr} exchange is required to make \glspl{sta} go from \textit{low-capability} to \textit{high-capability} mode before the transmission of \glspl{ppdu}. Similarly, for \glspl{sta} implementing a new mode called \gls{duo}---which allows them to inform their \gls{ap} about unavailable periods---an \gls{icf}-\gls{icr} exchange is needed to prepare scheduled \gls{duo} \glspl{sta} for participating in a coordinated \gls{txop}.
    \item \textbf{\gls{qos}:} \gls{mapc} Traffic Indication is a new functionality included as part of the \gls{mapc} operation, and serves to share information about traffic flows (i.e., \gls{qos} parameters including service start time and medium time) with other coordinated \glspl{ap}. This information can be very useful for deciding the set of \glspl{ap} that should be invited to a given coordinated \gls{txop}, which can be prioritized if they have latency-sensitive traffic, for example.
\end{itemize}

\section{Performance Evaluation of MAPC}
\label{sec:simulations}

\begin{table*}[t!]
    \centering
    \small  
    \begin{minipage}[t]{0.34\textwidth}
        \centering
        \caption{PHY layer parameters.}
        \resizebox{0.99\columnwidth}{!}{%
        \begin{tabular}{@{}llc@{}}
        \toprule
        \textbf{Param.} & \textbf{Descr.} & \textbf{Value} \\ \midrule
        $f_c$ & Center frequency & $5$ GHz \\
        $G_{tx}$ & TX gain & $0$ dB \\
        $G_{rx}$ & RX gain & $0$ dB \\
        $PL(d)$ & Path loss & \cite{merlin2015tgax} \\
        $N$ & Noise & $-95$ dBm \\
        $\sigma_{leg}$ & Leg. OFDM sym. duration & $4$ $\mu$s \\
        $\sigma$ & OFDM symbol duration & $16$ $\mu$s \\
        $GI$ & Guard interval & $3.2$ $\mu$s  \\
        $N_{ss}$ & Num. spatial streams & $1$ \\ \bottomrule
        \end{tabular}}
        \label{tab:phy_params}
    \end{minipage}
    \hfill 
    \begin{minipage}[t]{0.65\textwidth}
        \centering
        \caption{Frame and space durations and lengths.}
        \resizebox{0.9\columnwidth}{!}{%
        \begin{tabular}{@{}lp{6cm}c@{}}
        \toprule
        \textbf{Parameter} & \textbf{Description} & \textbf{Value} \\ \midrule
        $T_e$ & Empty slot duration & $9$ $\mu$s \\
        $T_\text{SIFS}$ & SIFS duration & $16$ $\mu$s \\
        $T_\text{AIFS}$ & AIFS duration & $34$ $\mu$s \\
        $T_\text{PHY-leg}$ & Legacy preamble duration & $20$ $\mu$s \\
        $T_\text{PHY-HE}$ & High-Efficiency preamble duration & $100$ $\mu$s \\
        $T_\text{BACK}$ & Block ACK duration & $32$ $\mu$s \\
        $T^{max}_\text{PPDU}$ & Maximum PPDU duration & $5.484$ ms \\        $L_\text{RTS}$/$L_\text{ICF}$/$L_{TF}$/$L_\text{MURTS}$ & Length of RTS, ICF, TF, and MU-RTS frames & $160$ bits \\
        $L_\text{CTS}$/$L_\text{ICR}$ & Length of CTS and ICR frames & $112$ bits \\
        $L_\text{MH}$ & Length of the MAC header & $320$ bits \\
        $L_\text{SF}$ & Length of the service field & $16$ bits \\ \bottomrule
        \end{tabular}}
        \label{tab:durations_lengths}
    \end{minipage}    
\end{table*}

This section presents a performance evaluation of the three \gls{mapc} schemes that operate on a per-\gls{txop} basis, i.e., \gls{cobf}, \gls{cosr}, and \gls{cotdma}. The goal is to showcase the major strengths and weaknesses of each mechanism, thus shedding light on their convenience in various situations and, in consequence, on their future beyond 802.11bn. For that, we purposely designed specific scenarios to show tradeoffs that are inherent to each \gls{mapc} feature. All evaluations were conducted using the \texttt{Kom8ndor} simulator~\cite{wilhelmi2026kom8ndor}, an 11bn-based extension of Komondor~\cite{barrachina2019komondor}.\footnote{The source code of \texttt{Kom8ndor}, under the GNU General Public License v3.0, is open and publicly available at \url{https://github.com/wn-upf/Komondor/tree/dev}, commit: cc289c4d75c2402d8934cc4155302af7d3d53a84.} By utilizing this open-source, unified tool, we maintain a consistent methodology across all experiments, ensuring that performance differences are the result of the schemes themselves. The main simulation parameters considered, common to all the experiments performed in the following subsections, are collected in Table~\ref{tab:phy_params} and Table~\ref{tab:durations_lengths}.

\subsection{Scenario 1 (Co-BF): Exhibition Hall}

For the realization of \gls{cobf}, we consider a simplified horizontal \gls{ula} model~\cite{balanis2016antenna} under a pure line-of-sight channel model, where the channel response is obtained from the steering angle and the corresponding \gls{ula} steering vector. This allows a clear interpretation of the results avoiding the increased complexity that multi-path would introduce. Specifically, each \gls{ap} computes a \gls{zf} precoding vector that steers a main beam toward its associated \gls{sta} while placing exact nulls toward the \glspl{sta} associated with the peer \glspl{ap}. The resulting beamforming gains are then applied to the received power, which is first computed using a simple path-loss model (see Table~\ref{tab:phy_params}). In addition, we assume that the steering angles are derived directly from the known node positions, which could be achieved with instantaneous and perfect \gls{csi}. Although the concept is demonstrated using this simplified channel model, the exact same principle extends to multipath environments when \gls{zf} is applied over the corresponding channel vectors \cite{YanGerQue2017}. In practice, obtaining such channel information may require feedback and channel estimation procedures; therefore, the study of imperfect and outdated feedback and the impact of channel estimation overheads remains a critical topic to explore (see, e.g.,~\cite{aboushehada2026autoencoder}).

\begin{figure}[t!]
    \centering
    \includegraphics[width=.85\columnwidth]{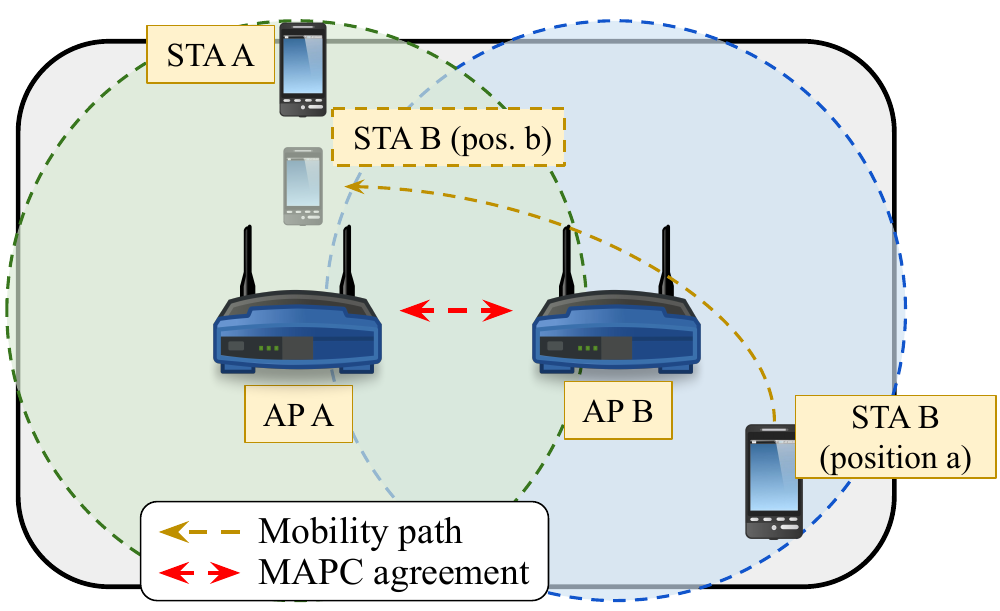}
    \caption{Scenario 1 (Co-BF): Exhibition hall.}
    \label{fig:scenario_bf}
\end{figure}

\textbf{Scenario description:} Two APs ($\text{AP}_A$, $\text{AP}_B$) serve two clients ($\text{STA}_A$, $\text{STA}_B$) who are \gls{xr} users and, in turn, require very high throughput and low latency communication. To show the strengths and weaknesses of \gls{cobf}, we consider two distinct situations in which $\text{STA}_B$ is located at different positions ($a$ and $b$), as shown in Fig.~\ref{fig:scenario_bf}. In position $a$, there is a large angular separation from the perspective of $\text{AP}_A$ ($\Delta\theta_A \approx 117^\circ$). In position $b$, in contrast, $\text{STA}_B$ is in line with $\text{STA}_A$ from $\text{AP}_A$'s perspective ($\Delta\theta_A = 0^\circ$), making it impossible to simultaneously apply a beam toward $\text{STA}_A$ while nulling $\text{STA}_B$ as we are considering pure line-of-sight channels. The simulation parameters used in this deployment are provided in Table~\ref{tab:cobf_parameters}.

\begin{table}[t!]
\caption{Simulation parameters for Sce. 1 (Co-BF).}
\label{tab:cobf_parameters}
\centering
\resizebox{.8\columnwidth}{!}{%
\begin{tabular}{@{}cc@{}}
\toprule
\textbf{Parameter} & \textbf{Value} \\ \midrule
AP locations (A, B) & $(3,0)$, $(6,0)$ m \\
STA (A) location & $(3,2)$ m \\
STA (B) location in pos. $a$ and $b$ & $(7,-2)$, $(3,1)$ m \\
Bandwidth, $B$ & $80$ MHz \\
Transmit power, $P$ & $20$ dBm \\
Number of antenna elements, $N_a$ & $8$ \\
Antenna spacing, $d$ & $0.5\lambda$ \\
Traffic load, $\rho$ & Full buffer\\ \bottomrule
\end{tabular}%
}
\end{table}

\begin{figure}[t!]
    \centering
    \subfloat[Throughput]{\includegraphics[width=\columnwidth]{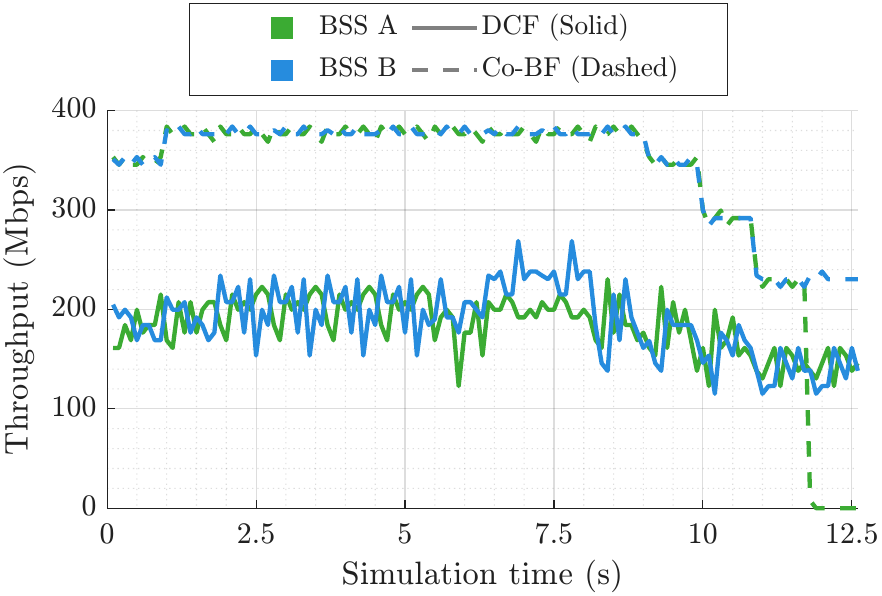}\label{fig:temp_throughput_cobf}}    
    \hfil
    \subfloat[Channel access delay]{\includegraphics[width=\columnwidth]{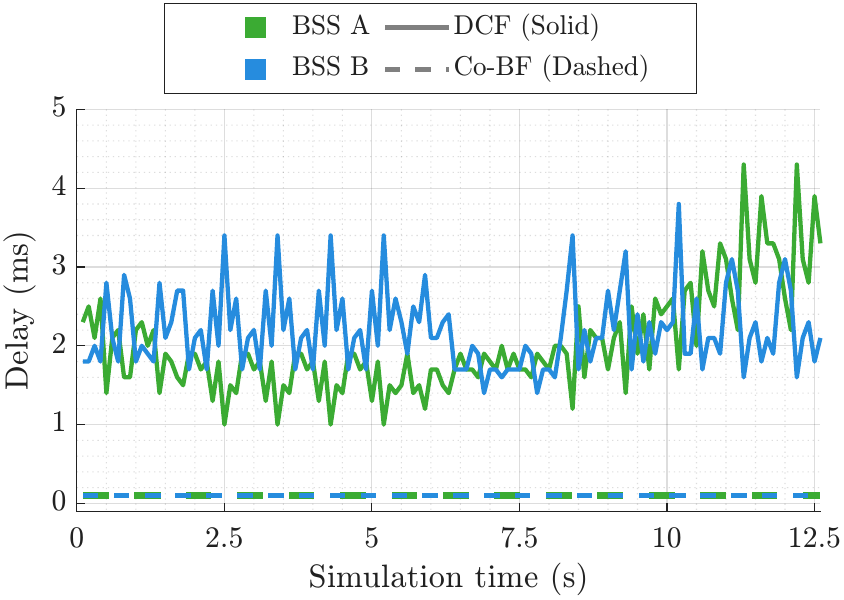}\label{fig:temp_delay_cobf}}
    \caption{Performance achieved by each BSS in Sce. 1 under DCF and Co-BF, from $\text{STA}_B$'s position $a$ to $b$.}
    \label{fig:performance_cobf}
\end{figure}

\textbf{Results (Fig.~\ref{fig:performance_cobf}):} We compare the performance of \gls{dcf} with \gls{cobf} as $\text{STA}_B$ moves across a range of positions, from position $a$ to $b$. The throughput achieved by each \gls{bss} over time is shown in Fig.~\ref{fig:temp_throughput_cobf}, where solid lines indicate the throughput under \gls{dcf} and dashed lines, under \gls{cobf}. Similarly, the channel access delay for both \glspl{bss} is shown in Fig.~\ref{fig:temp_delay_cobf}. Under \gls{dcf}, the performance remains stable throughout the simulation ($\sim$190-230~Mb/s), except for the 9th second of the simulation onwards, where the path loss between $\text{AP}_B$ and $\text{STA}_B$ increases as a result of the position of the latter, leading to lower effective data rates (and thus lower throughput, around 110~Mb/s) and higher channel access delay (above 3~ms). In those situations, the performance of $\text{BSS}_A$ is also affected due to the performance anomaly that stems from the shared channel in \gls{dcf}. Regarding \gls{cobf}, we observe three main situations:
\begin{itemize}
    \item \textit{Perfect interference suppression} (0-10~s): \gls{cobf} outperforms \gls{dcf} and provides ultra-high throughput ($\sim$375~Mb/s per \gls{bss}, a $\sim$90\% increase with respect to \gls{dcf}) because, for the locations of both \glspl{sta} during that period, the \gls{zf} precoder achieves exact interference suppression at the peer \gls{sta} while maintaining full receive power at the intended \gls{sta}. Such a perfect suppression provides a much more stable performance.
    \item \textit{Gradual degradation of \gls{zf}} (11-12~s):  As $\text{STA}_B$ approaches $(3,1)$~m, the angle between $\text{AP}_A$'s main beam and its null direction decreases progressively, hence reducing the performance of both \glspl{bss}, e.g., from 378~Mb/s at 10~s ($\Delta\theta_A=42^\circ$) to 229~Mb/s at 12~s ($\Delta\theta_A=13^\circ$).
    \item \textit{The \gls{zf} precoder collapses} (12-13~s): When reaching position $b$, $\text{STA}_B$ is placed between $\text{AP}_A$ and $\text{STA}_A$ ($\Delta\theta_A=0^\circ$), making it impossible for $\text{AP}_A$ to simultaneously place both a beam and a null toward the same direction. This makes $\text{AP}_A$'s precoder collapse, so $\text{BSS}_A$ achieves 0~Mb/s. In that position, $\text{AP}_B$ can still transmit successfully to $\text{STA}_B$ ($\Delta\theta_B=15^\circ$), achieving $\sim$230~Mb/s.   
\end{itemize}

Regarding channel access delay (shown in Fig.~\ref{fig:temp_delay_cobf}), \gls{cobf} provides near-instantaneous channel access ($<$0.1~ms) to both \glspl{ap} throughout all evaluated positions, due to the coordinated nature of the protocol and the joint participation of both \glspl{bss} in all \gls{txop}. However, as shown by the throughput results, \gls{cobf} can lead to near-zero performance when beam alignment is geometrically infeasible. Regarding \gls{dcf}, it exhibits a varying channel access delay of up to $\sim$4~ms that tends to increase as the channel conditions worsen. 

\textbf{Takeaways \gls{cobf}:} The effectiveness of \gls{cobf} depends on the angular separation between devices and the degrees of freedom available to the \gls{ap} antenna array. In scenarios with high spatial diversity (Sce. 1, position $a$), \gls{cobf} provides a huge boost of aggregate throughput compared to \gls{dcf}, unlocking next-generation applications like \gls{xr} in dense and small enclosures. However, in scenarios with poor angular separation (Sce. 1, position $b$), the proximity of the nodes may cause both steering and nulling vectors to overlap, resulting in poor \gls{sinr} and the loss of throughput, which in turn leads to decreased reliability due to frequent packet losses. The main takeaways of \gls{cobf} are summarized in Table~\ref{tab:takeaways_cobf}.

\begin{table}[ht!]
\centering
\caption{Summary of the performance advantages ($\checkmark$) and disadvantages (\ding{55}) of Co-BF (compared to DCF) regarding throughput, delay, and reliability.}
\label{tab:takeaways_cobf}
\resizebox{.8\columnwidth}{!}{%
\begin{threeparttable}
\begin{tabular}{@{}lccc@{}}
\toprule
 & \textbf{$\uparrow$ Throughput} & \textbf{$\downarrow$ Delay} & \textbf{$\uparrow$ Reliability} \\ \midrule
\textbf{Position $a$} & $\checkmark$ & $\checkmark$ & $\checkmark$ \\ 
\textbf{Position $b$} & \ding{55} & $\checkmark$\tnote{*} & \ding{55} \\ \bottomrule
\end{tabular}
\begin{tablenotes}
\scriptsize
\item $^*$Despite several transmissions attempted by AP$_\text{A}$ are unsuccessful when STA$_\text{B}$ is in position $b$, the channel access delay is still low.
\end{tablenotes}
\end{threeparttable}
} 
\end{table}

\subsection{Scenario 2 (Co-SR): Dual Apartment}

\begin{figure}[t!]
    \centering
    \includegraphics[width=\columnwidth]{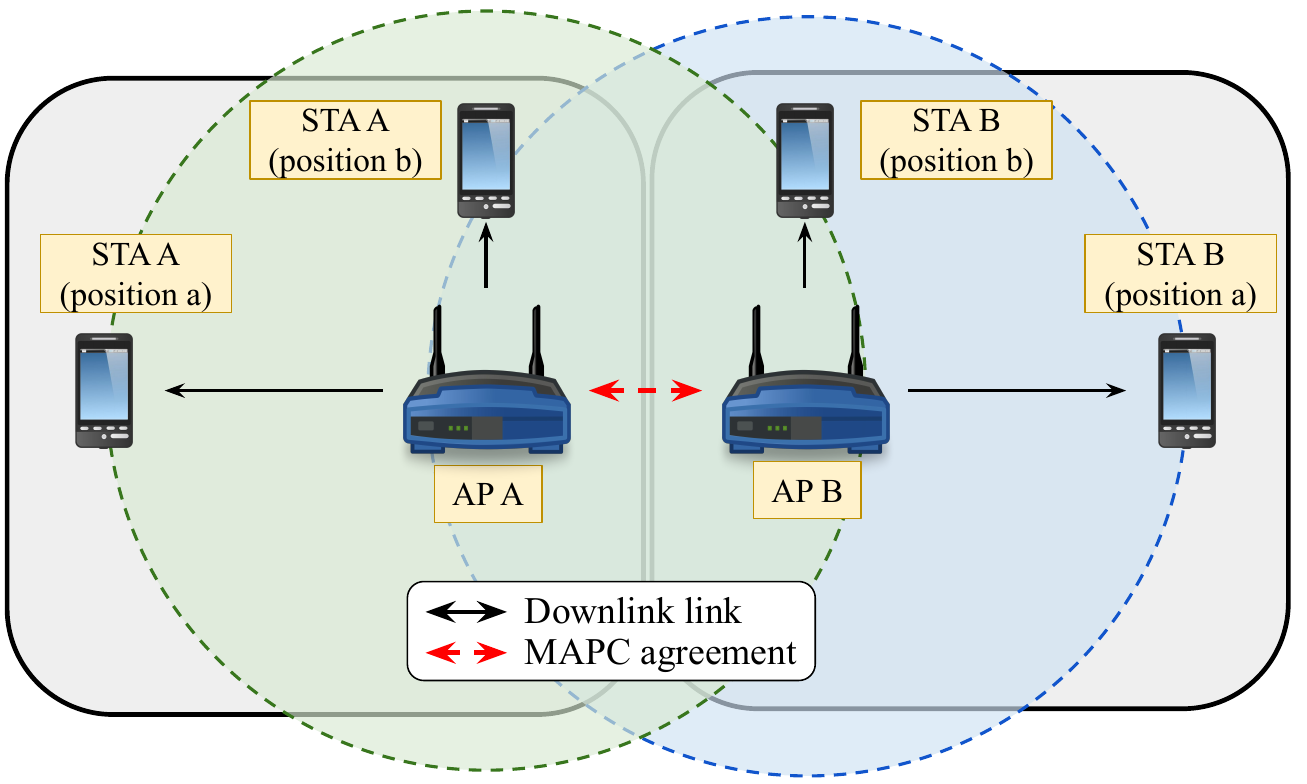}
    \caption{Scenario 2 (Co-SR): Dual residential apartment.
    }
    \label{fig:scenario_sr}
\end{figure}

\textbf{Scenario description (Fig.~\ref{fig:scenario_sr}):} Two adjacent apartments with one \gls{ap} ($\text{AP}_A$ and $\text{AP}_B$) and one \gls{sta} ($\text{STA}_A$ and $\text{STA}_B$) each. Although the scenario places $\text{AP}_A$
$\text{AP}_B$ in adjacent apartments separated by a wall, the position of the wall is not geometrically resolved. In particular, the TGax residential path loss model~\cite{merlin2015tgax} adopted here applies a fixed wall penetration loss term derived from the average building geometry of that scenario, which captures the average effect of inter-apartment walls. Two deployments (Deployment 1 and Deployment 2), in which \glspl{sta} are in different locations, are considered to illustrate the strengths and weaknesses of \gls{cosr}. In Deployment 1, both \glspl{sta} are located far away from the other \gls{bss}, making space highly reusable. In Deployment 2, instead, \glspl{sta} are placed near the other \gls{ap}, making it hard to reuse space. As for the transmit power used by the coordinated \gls{ap} in shared \glspl{txop}, it is limited by the coordinating \gls{ap} to 20~dBm and 15~dBm in Deployments 1 and 2, respectively. The main simulation parameters used in Scenario 2 are shown in Table~\ref{tab:cosr_parameters}. Transmissions are omnidirectional.

\begin{table}[t!]
\caption{Simulation parameters for Sce. 2 (Co-SR).}
\label{tab:cosr_parameters}
\centering
\resizebox{.8\columnwidth}{!}{%
\begin{tabular}{@{}cc@{}}
\toprule
\textbf{Parameter} & \textbf{Value} \\ \midrule
AP locations (A, B) & $(3,0)$, $(5,0)$ m \\
STA location (A, B) in Depl. 1 & $(1,0)$, $(7,0)$ m \\
STA location (A, B) in Depl. 2 & $(3,2)$, $(5,2)$ m \\
Bandwidth, $B$ & $20$ MHz \\
Transmit power, $P$ & $20$ dBm \\
Co-SR transmit power (Depl. 1/2) & $20$/$15$ dBm \\
Traffic load, $\rho$ & $[1000-5000]$ pkt/s\\ \bottomrule
\end{tabular}%
}
\end{table}

\begin{figure}[t!]
    \centering
    \subfloat[Sce. 2, Deployment 1]{\includegraphics[width=0.85\columnwidth]{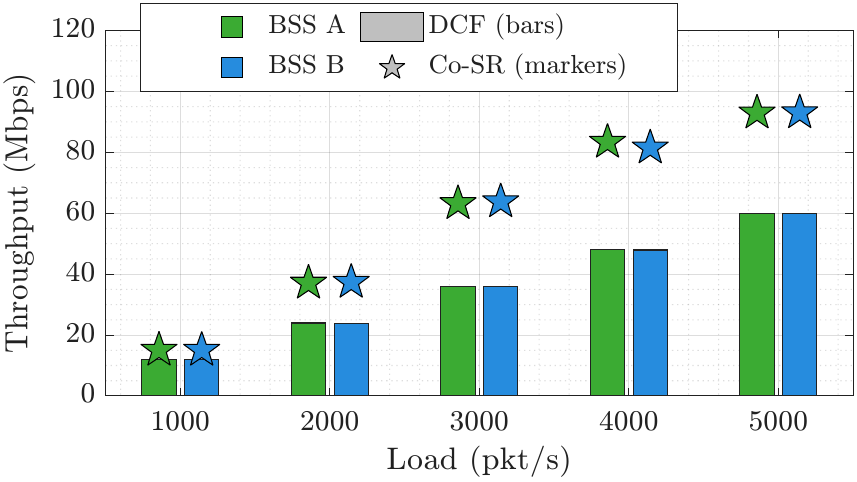}\label{fig:throughput_cosr_deployment1}}    
    \hfil
    \subfloat[Sce. 2, Deployment 2]{\includegraphics[width=0.85\columnwidth]{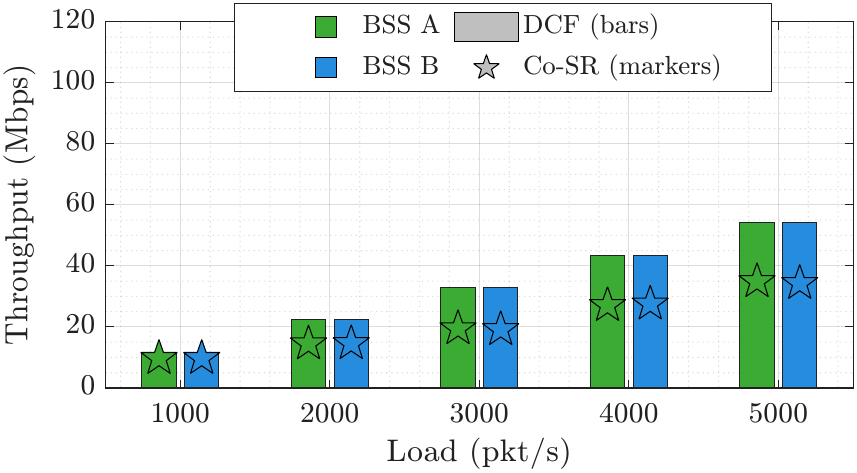}\label{fig:throughput_cosr_deployment2}}
    \caption{Mean throughput achieved by each BSS under DCF and Co-SR in Depl. 1 and 2, for $\rho \in \{1000, 2000, 3000, 4000, 5000\}$~pkt/s.}
    \label{fig:throughput_cosr}
\end{figure}

\begin{figure}[ht!]
    \centering
    \includegraphics[width=\columnwidth]{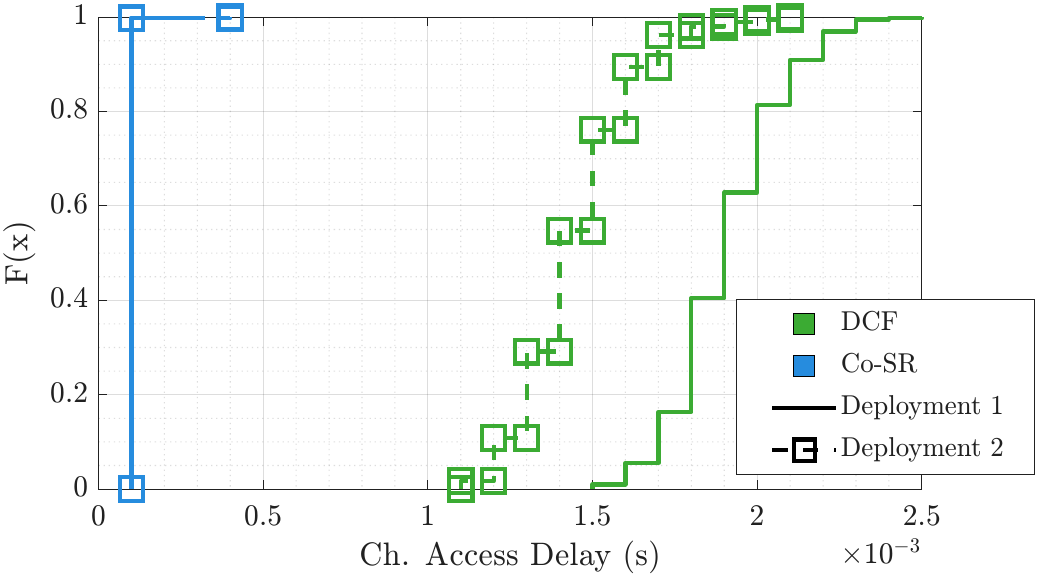}
    \caption{CDF of the channel access delay experienced across the two BSSs in Sce. 2 under DCF and Co-SR, for Depl. 1 and 2.}
    \label{fig:cdf_delay_cosr}
\end{figure}

\textbf{Results (Fig.~\ref{fig:throughput_cosr} and Fig.~\ref{fig:cdf_delay_cosr}):} We study the performance of \gls{cosr} compared to \gls{dcf} 
 in terms of throughput and channel access delay. We consider two settings (Deployment 1 and 2) that illustrate situations that favor and discourage the utilization of \gls{cosr}. 
 \begin{itemize}
     \item \textit{Deployment 1}: As shown in Fig.~\ref{fig:throughput_cosr_deployment1}, the throughput of both \glspl{bss} increases equally with the traffic load (from 1000 to 5000 pkt/s). \gls{cosr} (star markers) consistently improves \gls{dcf} for all the traffic load values. In addition, as the load increases, the performance gap between the two mechanisms widens significantly, underscoring the efficiency of \gls{cosr} under higher loads. When it comes to the overall channel access delay (Fig.~\ref{fig:cdf_delay_cosr}), \gls{cosr} clearly outperforms \gls{dcf} ($\sim$95\% decrease) for both \glspl{bss} to transmit during shared \glspl{txop}. Only in a few cases, the access delay is not instantaneous, leading to spikes of 0.4~ms due to collisions by backoff (both \glspl{ap} attempt to transmit an \gls{icr} frame at the same time).
     \item \textit{Deployment 2}: By looking at the throughput (Fig.~\ref{fig:throughput_cosr_deployment2}), we observe that the gap between \gls{dcf} and \gls{cosr} increases with the load, but this time in favor of \gls{dcf}. At the highest load (5000 pkt/s), \gls{dcf} reaches $\sim$55~Mb/s because both \glspl{bss} alternate the medium. In turn, \gls{cosr} leads to $\sim$35~Mb/s for each \gls{bss}, which is due to the packet losses suffered by the coordinated \gls{ap} in such a high interference situation. 
     Notice that the coordinating \gls{bss} can still perform successful transmissions during shared \glspl{txop}, provided that it is the coordinated \gls{ap} which needs to lower its transmit power. Since both \glspl{ap} alternate the role of coordinator, they end up obtaining a very similar performance. In terms of channel access delay, the behavior is the same as for Deployment 1, being \gls{cosr} faster than \gls{dcf}, but this time leading to unsuccessful transmissions.
 \end{itemize}

\textbf{Takeaways \gls{cosr}:} The results glimpse that the effectiveness of \gls{cosr} is heavily dependent on the location of the devices and the resulting interference to \glspl{sta}. In favorable spatial reuse conditions (Deployment 1), \gls{cosr} is superior to \gls{dcf} in terms of aggregate throughput, scaling efficiently and reliably with the traffic load. Furthermore, it offers near-instantaneous channel access. In contrast, in high-interference settings (Deployment 2), while \gls{cosr} maintains very low and stable channel access delays, it penalizes throughput and reliability due to packet losses suffered by the coordinated \glspl{ap}. The main takeaways of \gls{cosr} are summarized in Table~\ref{tab:takeaways_cosr}.

\begin{table}[ht!]
\caption{Summary of the performance advantages ($\checkmark$) and disadvantages (\ding{55}) of Co-SR (compared to DCF) regarding throughput, delay, and reliability.}
\label{tab:takeaways_cosr}
\centering
\resizebox{.8\columnwidth}{!}{%
\begin{tabular}{@{}lccc@{}}
\toprule
 & \textbf{$\uparrow$ Throughput} & \textbf{$\downarrow$ Delay} & \textbf{$\uparrow$ Reliability} \\ \midrule
\textbf{Deployment 1} & $\checkmark$ & $\checkmark$ & $=$ \\
\textbf{Deployment 2} & \ding{55} & $\checkmark$ & \ding{55}\\ \bottomrule
\end{tabular}%
}
\end{table}

\subsection{Scenario 3 (Co-TDMA): Shared Workspace in Enterprise}

\begin{figure}[ht!]
    \centering
    \includegraphics[width=\columnwidth]{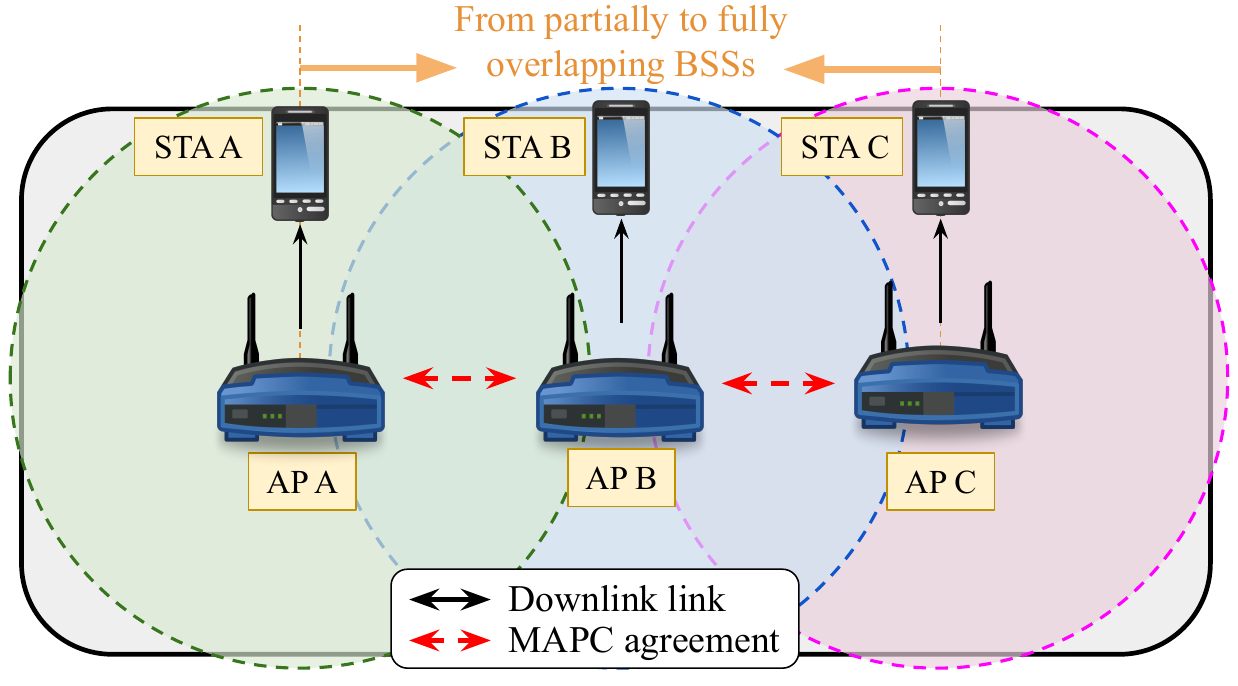}
    \caption{Scenario 3 (Co-TDMA): Shared workspace in enterprise.}
    \label{fig:scenario_cotdma}
\end{figure}

\textbf{Scenario description (Fig.~\ref{fig:scenario_cotdma}):} A long rectangular office floor with three APs ($\text{AP}_A, \text{AP}_B, \text{AP}_C$) placed in a line, each generating 60~Mb/s traffic with Poisson arrivals. Two settings are considered. In full overlap, all the \glspl{ap} listen to each other. In partial overlap, $\text{AP}_A$ and $\text{AP}_C$ are sufficiently far apart not to interfere with each other, but $\text{AP}_B$ (between $\text{AP}_A$ and $\text{AP}_C$) is within the interference range of both (\emph{flow-in-the-middle starvation}). Regarding the \gls{cotdma} configuration, we consider two cases: \gls{cotdma} (I), where $\text{AP}_A$ and $\text{AP}_B$ form one group (\{$A$, $B$\}), and \gls{cotdma} (II), where $\text{AP}_B$ forms two groups with $\text{AP}_A$ and $\text{AP}_C$, respectively (\{$A$, $B$\} and \{$B$, $C$\}). In \gls{cotdma} (II), $\text{AP}_B$ alternates \gls{txop} sharing with the other two \glspl{ap} using Round-Robin, while $\text{AP}_A$ and $\text{AP}_C$ always share their acquired \glspl{txop} with $\text{AP}_B$. The main simulation parameters for this scenario are collected in Table~\ref{tab:cotdma_parameters}.

\begin{table}[h!]
\caption{Simulation parameters for Sce. 3 (Co-TDMA).}
\label{tab:cotdma_parameters}
\centering
\resizebox{0.95\columnwidth}{!}{%
\begin{tabular}{@{}cc@{}}
\toprule
\textbf{Parameter} & \textbf{Value} \\ \midrule
AP locations (A, B, C) & $(0,0)$, $(2,0)$, $(4,0)$ m \\
STA location (A, B, C) & $(0,3)$, $(2,3)$, $(4,3)$ m \\
Bandwidth, $B$ & $40$ MHz \\
Transmit power, $P$ & $15$ dBm \\ 
Co-TDMA TXOP sharing policy & Equal split \\
Traffic load, $\rho$ & $5000$ pkt/s\\ \bottomrule
\end{tabular}%
}
\end{table}

\begin{figure}[t!]
    \centering
    \subfloat[Sce. 3, Full overlap]{\includegraphics[width=.8\columnwidth]{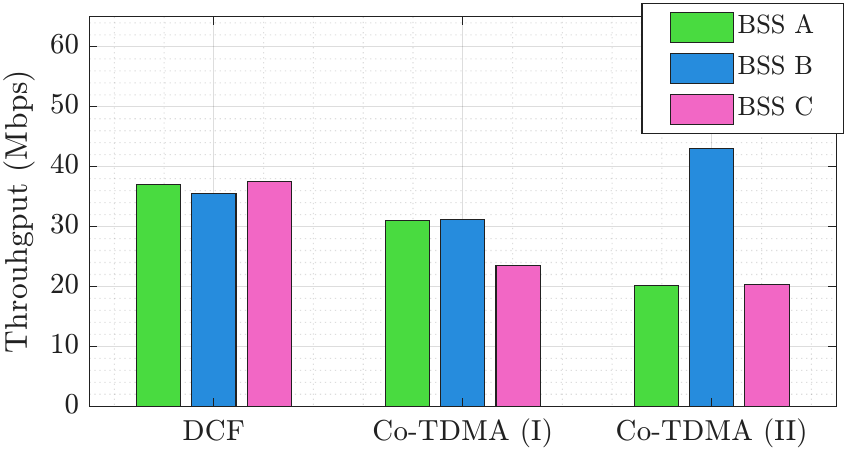}\label{fig:throughput_cotdma_full_overlap}}    
    \hfil
    \subfloat[Sce. 3, Partial overlap]{\includegraphics[width=.8\columnwidth]{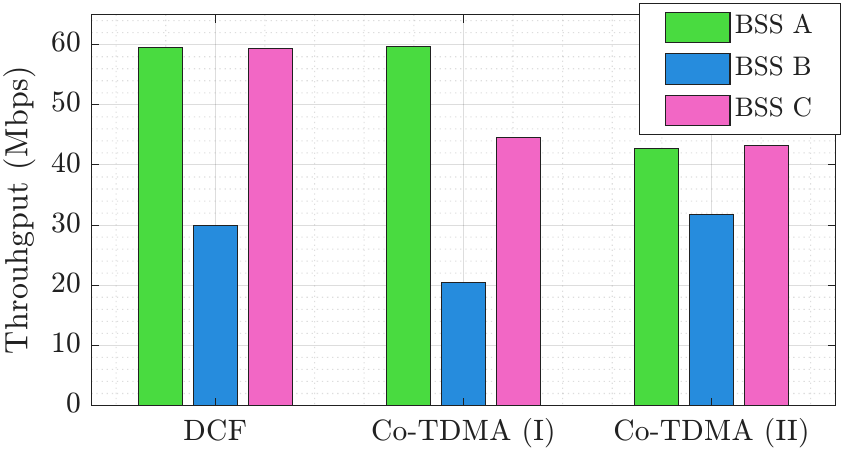}\label{fig:throughput_cotdma_partial_overlap}}
    \caption{Mean throughput achieved by each BSS in Sce. 3 under DCF, Co-TDMA (I), and Co-TDMA (II).
    }
    \label{fig:throughput_cotdma}
\end{figure}

\begin{figure}[t!]
    \centering
    \subfloat[Full overlap]{\includegraphics[width=\columnwidth]{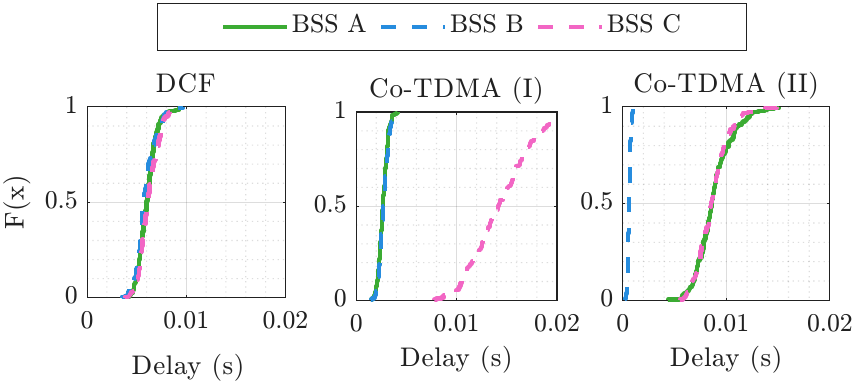}\label{fig:avg_delay_cotdma_full_overlap}}   
    \hfil
    \subfloat[Partial overlap]{\includegraphics[width=\columnwidth]{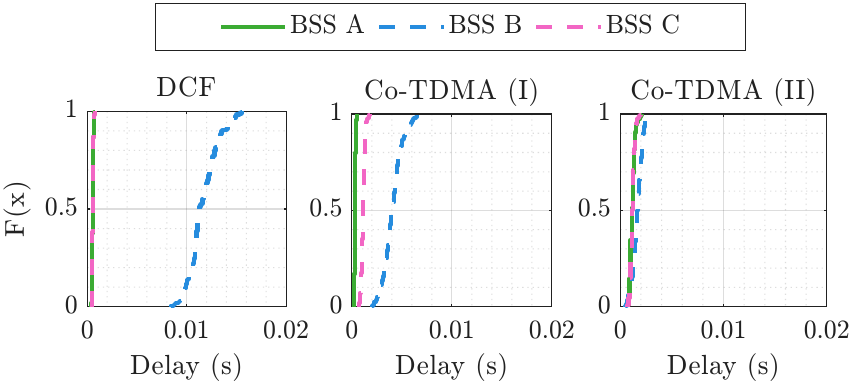}\label{fig:avg_delay_cotdma_partial_overlap}}
    \caption{CDF of the channel access delay experienced by each BSS under DCF, Co-TDMA (I), and Co-TDMA (II).}
    \label{fig:cdf_delay_cotdma}
\end{figure}

\textbf{Results (Fig.~\ref{fig:throughput_cotdma} and Fig.~\ref{fig:cdf_delay_cotdma}):} The performance of \gls{cotdma} in terms of average \gls{bss} throughput and channel access delay is compared to \gls{dcf}, for both full overlap and partial overlap settings. The two settings reveal complementary insights: \gls{cotdma} is valuable where \gls{dcf} struggles (e.g., unfair topologies, latency-sensitive traffic), but it offers little advantage where \gls{dcf} already performs well.
\begin{itemize}
    \item \textit{Full overlap}: This setting is favorable to \gls{dcf}, which achieves equal throughput across all \glspl{bss} since the channel is shared evenly (Fig.~\ref{fig:throughput_cotdma_full_overlap}). For \gls{cotdma}, aggregate throughput decreases due to the random nature of packet arrivals (following a Poisson process), making shared \gls{txop} less efficient than opportunistically accessing the channel via \gls{dcf} (the coordinated \gls{ap} does not always have enough buffered traffic to fill its allocated time). The benefit of \gls{cotdma} in this setting lies in the channel access delay (Fig.~\ref{fig:avg_delay_cotdma_full_overlap}). \gls{cotdma} (I) reduces the delay for the coordinated BSSs ($A$ and $B$) by 55\% compared to \gls{dcf}, providing the predictable, low-latency access that time-sensitive applications require. This gain, however, comes at a cost for the non-coordinated \gls{bss} ($C$), whose delay worsens by more than 130\%. When it comes to \gls{cotdma} (II), \gls{bss} $B$ (which is a member of both groups) obtains more resources than the rest because it participates in all the \glspl{txop}.
    \item \textit{Partial overlap}: This setting exposes the weaknesses of \gls{dcf} (flow-in-the-middle starvation of \gls{bss} $B$), and is where \gls{cotdma} demonstrates its true value. In terms of throughput (Fig.~\ref{fig:throughput_cotdma_partial_overlap}), \gls{cotdma} (I) not only does not solve the issue, but also lowers the aggregate performance as a result of the inefficiencies with Poisson traffic highlighted above. \gls{cotdma} (II) allows softening the starvation problem by slightly improving the throughput of \gls{bss} $B$, yet it does not fully solve the issue. Nevertheless, the most significant gains of \gls{cotdma} are in terms of channel access delay (Fig.~\ref{fig:avg_delay_cotdma_partial_overlap}). As shown, \gls{cotdma} (I) reduces it by $\sim$65\% with respect to \gls{dcf}, and \gls{cotdma} (II) goes further, equalizing the access delay across all \glspl{bss} at a low, stable value that outperforms \gls{dcf} by $\sim$70\%.
\end{itemize}

\textbf{Takeaways \gls{cotdma}:} The results show that the benefits of \gls{cotdma} are highly dependent on the network topology and traffic profile. \gls{cotdma} demonstrates its clearest value in topologies where \gls{dcf} is inherently unfair. In the partially overlapping deployment, where \gls{dcf} leads to flow-in-the-middle starvation, an appropriate \gls{cotdma} grouping strategy simultaneously improves latency, stability, and fairness, in line with \gls{uhr} targets. Advanced coordination like \gls{cotdma} (II) excels in this regard, keeping the channel access delay low and stable across \textit{all} \glspl{bss}. Similarly, for any \gls{bss} with low-latency traffic requirements, belonging to multiple \gls{mapc} groups is beneficial to meet its goals, regardless of the topology. 

By contrast, \gls{cotdma}'s limitations become evident in settings where \gls{dcf} already performs well. In the fully overlapping setup, \gls{dcf} is superior in terms of aggregate throughput and system-wide fairness because opportunistic channel access under stochastic traffic proves to be more efficient. Moreover, \gls{cotdma} is shown to prioritize coordinated nodes, degrading overall efficiency and heavily penalizing legacy devices. This suggests that, in such scenarios, \gls{cotdma} should be activated and configured intelligently rather than applied broadly. Note that, although our analysis focused on downlink \gls{udp}-like traffic, \gls{cotdma}'s deterministic nature has also been shown in the literature to be suitable for periodic traffic patterns and uplink transmissions, enabling better support for closed-loop protocols such as \gls{tcp}~\cite{val2025wi}. The main takeaways of \gls{cotdma} are summarized in Table~\ref{tab:takeaways_cotdma}.

\begin{table}[ht!]
\caption{Summary of the performance advantages ($\checkmark$) and disadvantages (\ding{55}) of Co-TDMA (compared to DCF) for legacy and Co-TDMA devices.}
\label{tab:takeaways_cotdma}
\resizebox{\columnwidth}{!}{%
\begin{tabular}{@{}lccccc@{}}
\toprule
\multicolumn{1}{c}{\textbf{}} & \multicolumn{2}{c}{\textbf{$\uparrow$ Throughput}} & \multicolumn{2}{c}{\textbf{$\downarrow$  Latency}} & \multicolumn{1}{c}{\textbf{$\uparrow$ Fairness}} \\ \midrule
 & \multicolumn{1}{c}{\textit{Legacy}} & \multicolumn{1}{c}{\textit{Co-TDMA}} & \multicolumn{1}{c}{\textit{Legacy}} & \multicolumn{1}{c}{\textit{Co-TDMA}} & \multicolumn{1}{c}{\textit{All}} \\
\textbf{Full overlap} & \ding{55} & \ding{55} & \ding{55} & $\checkmark$ & \ding{55} \\
\textbf{Partial overlap} & \ding{55} & \ding{55} & \ding{55} & $\checkmark$ & $\checkmark$ \\ \bottomrule
\end{tabular}%
}
\end{table}

\section{The Future of MAPC}
\label{sec:future_mapc}

In the evolving landscape of Wi-Fi, particularly with the advent of 11be and the upcoming 11bn, numerous contributions have been made to shape the future 802.11 amendments. Beyond the overviewed \gls{mapc} features described above, other emerging mechanisms have also been broadly discussed in standardization: \gls{jt}, \gls{coofdma}, and \gls{conpca}. While those features will not be part of 11bn, they can potentially be introduced later, e.g., in 802.11bx (\mbox{Wi-Fi} 9)~\cite{karamyshev2025towards}. In addition, other functionalities that build upon the 11bn \gls{mapc} framework may emerge.

\subsection{Extension of Current MAPC Schemes}

The 11bn \gls{mapc} framework considers a coordinated operation of up to two \glspl{ap}, but it has already been shown that coordinating more \glspl{ap} or blending multiple coordination schemes~\cite{tcl2025mapc} could be a reasonable target for Wi-Fi 9. In this paper (Section~\ref{sec:simulations}), we also showed the potential and the complexity of extending the coordination to multiple \glspl{ap}, including illustrative examples where $\geq 2$-\gls{ap} coordination either succeeds or fails in achieving certain performance goals such as aggregate performance or fairness. To this, one must add the exponential coordination overheads (e.g., in \gls{cobf}, overheads are already significant for two \glspl{ap}~\cite{you2025improving}), the complexity of exponential resource allocation (e.g., allocating \glspl{ru} among multiple \glspl{ap} and \glspl{sta} is NP-hard~\cite{parizi2025resource}), or implementation aspects such as synchronization. Similarly, extending the 11bn framework to seamlessly work with \gls{mlo} provides as many opportunities as challenges it poses~\cite{ofinno2025mlo, zhang2023ieee, kumar2025multi}. 

\subsection{Joint Transmissions (JT)}

\gls{jt}, or distributed \gls{mu}-\gls{mimo}, is a multi-device multi-antenna technique where multiple \glspl{ap} transmit the same data stream to a single or multiple \glspl{sta} at the same time, to provide a significantly higher \gls{sinr} to them~\cite{lu2014overview}. With \gls{jt}, an interfering AP becomes a server, converting interference into a useful signal. Provided that the signals from the different \glspl{ap} follow different paths to reach the user (hence they naturally arrive at a different phase), \gls{jt} considers coherent beamforming to compensate for that and make the different transmissions constructive at the receiver side~\cite{nigam2014coordinated}. However, to perform \gls{jt} in practice, a tight time synchronization among the different transmitters is required. In addition, the involved \glspl{ap} would have to exchange not only the \gls{csi} feedback to compute the necessary precoding weights, but also the data to be transmitted to the user, hence requiring a dedicated high-speed backhaul link. Despite this, \gls{jt} has been shown to be very effective at improving reliability and throughput~\cite{kim2024multi}.

\subsection{Coordinated OFDMA (Co-OFDMA)}

\gls{ofdma} was introduced in 11ax as a  multi-user version of \gls{ofdm}, enabling the subdivision of channel resources into \glspl{ru} of different sizes and assigning them to multiple \glspl{sta}. While \gls{ofdma} is useful to coordinate users within a single \gls{bss}, it faces limitations in \gls{obss} deployments, where the availability of \glspl{ru} is not guaranteed. In this regard, \gls{coofdma} would allow allocating \gls{ofdma} channel resources wisely across multiple coordinated \glspl{bss}, thereby mitigating inter-\gls{bss} interference, lowering the worst-case latency, and fundamentally improving reliability. Initially, \gls{coofdma} could be used in the context of \gls{txop} sharing, with a coordinating \gls{ap} assigning idle \glspl{ru} to other coordinated \glspl{ap}~\cite{samsung2020ofdma}. Alternatively, \gls{coofdma} could allow determining the status of certain \glspl{ru} (e.g., an \gls{ap} can avoid assigning reused \gls{ru} to cell-edge users, where interference is higher) or even be combined with other features like \gls{cobf} or \gls{cosr}. Some works that have studied and anticipated the performance gains of \gls{coofdma} are~\cite{huawei2019ofdma, imputato2024meeting, hinohara2024multiple, parizi2025resource}. However, the feasibility of \gls{coofdma} is hindered by its stringent synchronization requirements. Unlike intra-\gls{bss} \gls{ofdma}, where a single \gls{ap} maintains a clock for the associated \glspl{sta}, \gls{coofdma} requires synchronization between multiple independent \glspl{ap}. In this regard, the lack of synchronization and alignment between the oscillators from the multiple \glspl{ap} in \gls{coofdma} would result in inter-carrier interference and, as a result, performance degradation.

\subsection{Coordinated Non-Primary Channel Access (Co-NPCA)}

\gls{npca} is a new 11bn feature that was introduced to provide greater flexibility in accessing the spectrum. In particular, \gls{npca}-compliant devices can temporarily transfer their \gls{bss} operation to a different channel (named \gls{npca} primary channel) and potentially transmit there when they sense their primary channel busy. This has been shown to be particularly useful in \gls{obss} scenarios, provided that higher transmission opportunities are created~\cite{bellalta2025modelling, lopez2025wi}. To further improve \gls{npca}, several standard contributions have already introduced \gls{conpca}, a technique whereby multiple \glspl{ap} agree upon the definition of \gls{npca} primary channels~\cite{zte2024npca, ofinno2025npca, huawei2024npca}. This is expected to boost reliability in challenging \gls{obss} scenarios.

\subsection{Other Potential Coordinated Features}

\gls{mapc} significantly expands the way current Wi-Fi networks operate, opening new avenues for architectural improvements and innovation. Beyond \gls{phy} and \gls{mac} layer enhancements, \gls{mapc} could potentially enable new features, including:
\begin{itemize}
    \item \textbf{Advanced channel planning:} In line with what \gls{cocr} proposes, different \glspl{bss} could coordinate and optimally select operating channels, thereby mitigating co-channel interference and enhancing spatial reuse~\cite{balbi2012centralized}.
    \item \textbf{Seamless roaming:} With \gls{mapc}, seamless handovers can be provided, avoiding interruptions and improving reliability. In \cite{kim2024multi}, a \gls{jt} approach is proposed for unlocking roaming across multiple \glspl{ap}.
    \item \textbf{Energy efficiency:} Coordinated \glspl{ap} could dynamically adjust their configurations (e.g., transmit power) for the sake of saving energy~\cite{tsao2011survey}. Furthermore, coordination could be leveraged to optimize energy efficiency by intelligently distributing the users' load across multiple \glspl{ap} and guiding devices to select the most energy-efficient \gls{ap}~\cite{wu2015energy}.
    \item \textbf{Advanced sensing and localization services:} By combining signal measurements from multiple \glspl{ap} or applying coordinated power control, Wi-Fi could achieve highly accurate indoor positioning and sensing capabilities~\cite{meneghello2023toward, huang2022coordinated}.
    \item \textbf{Enhanced security:} Coordinated \glspl{ap} could share information about suspicious traffic patterns or anomalous device behavior, allowing for faster and more effective detection and mitigation of threats~\cite{takahashi2010ieee, maurice2013improving}. \gls{mapc} could also allow the detection and mitigation of jamming attacks (e.g., by employing coordinated nulling towards the jammer). Similarly, privacy could be enhanced.
    \item \textbf{AI/ML empowerment:} \gls{mapc} could also potentially enable joint \gls{ai}/\gls{ml} operations, including distributed model training or reward sharing~\cite{wilhelmi2024machine, wilhelmi2024coordinated}. In the case of \gls{cobf}, which involves a lot of sounding signaling, a lot of attention has been placed in \gls{ai} methods such as autoencoders for compressing \gls{csi} feedback~\cite{aboushehada2026autoencoder}. Apart from that, \gls{ai}-based joint schedulers have been proposed as a compelling solution for \gls{mapc}~\cite{nunez2025deep}.
\end{itemize}

\section{Conclusions}
\label{sec:conclusions}

\gls{mapc} is set to be the next major milestone in Wi-Fi, enabling, for the first time, seamless cooperation between \glspl{ap}. The 802.11bn framework introduces five core coordination schemes (\gls{cobf}, \gls{cosr}, \gls{cotdma}, \gls{cortwt}, and \gls{cocr}) designed to transform autonomous channel access and network operation into a cooperative setup. However, the major disruption behind \gls{mapc} is yet to come, as its 11bn framework serves as a cornerstone for future amendments beyond ultra-high reliability. In this paper, we have overviewed the 11bn \gls{mapc} framework as defined in D1.4 and described its considered coordination schemes, pinpointing their key strengths and weaknesses through simulation results. Finally, we have discussed the future of \gls{mapc}, including other schemes under discussion and potential ways forward.

\section*{Acknowledgment}

This paper is supported by TRUE Wi-Fi PID2024-155470NB-I00 (MICIU/AEI/10.13039/501100011033/FEDER, UE), MLDR (Chist-ERA WAI 2022) PCI2023-145958-2 MCIU/AEI/10.13039, AGAUR ICREA Academia 00077, by PID2021-123999OB-I00 and PID2024-156488OB-I00 (MICIU/AEI/FEDER,UE), by CEX2021-001195-M/AEI/10.13039/501100011033 and CNS2023-145384,
by CIDEGENT PlanGenT, Grant CIDEXG/2022/17, Project iTENTE, and 
by CNS2023-144333 (MCIN/AEI/10.13039/501100011033).

\appendix[Elements, Messages, and Frame Formats]
\label{sec:appendix_a}

\subsection{MAPC Element}
\label{app:mapc_element}

\begin{figure*}[ht]
    \centering
    \includegraphics[width=.8\textwidth]{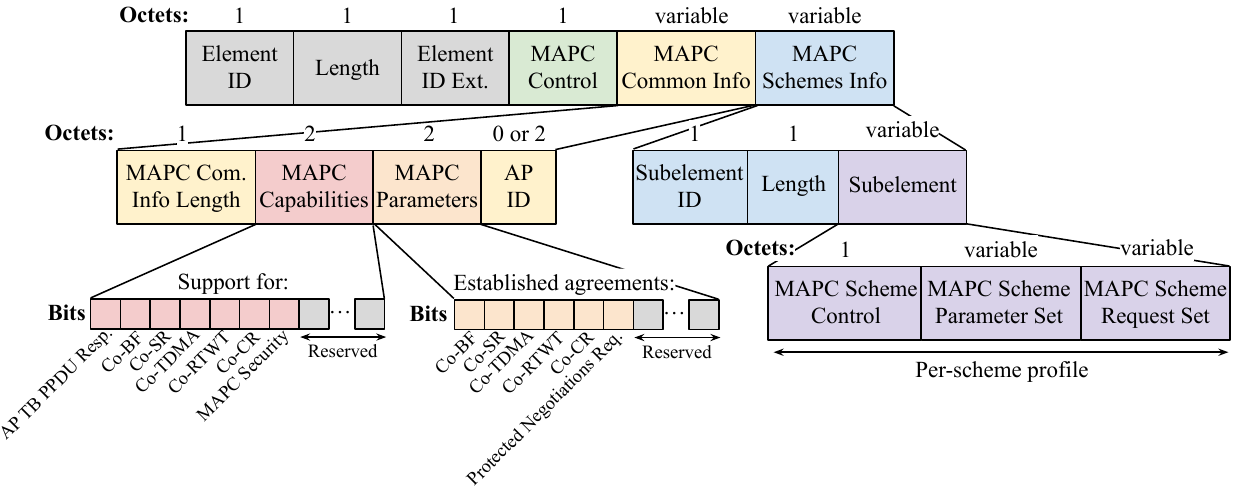}
    \caption{New \gls{mapc} element used to advertise an AP's supported coordination schemes, ongoing agreement status, and per-scheme parameters, enabling the discovery and negotiation of MAPC agreements between APs.}
    \label{fig:mapc_element}
\end{figure*}

\begin{figure}[ht]
\centering
\subfloat[]{\includegraphics[width=0.45\textwidth]{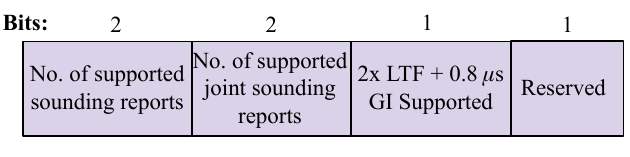}%
\label{fig:cobf_profile}}
\vfil
\subfloat[]{\includegraphics[width=0.23\textwidth]{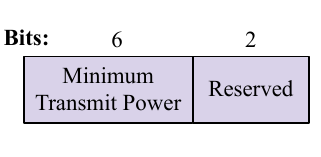}%
\label{fig:cosr_profile}}
\vfil
\subfloat[]{\includegraphics[width=0.48\textwidth]{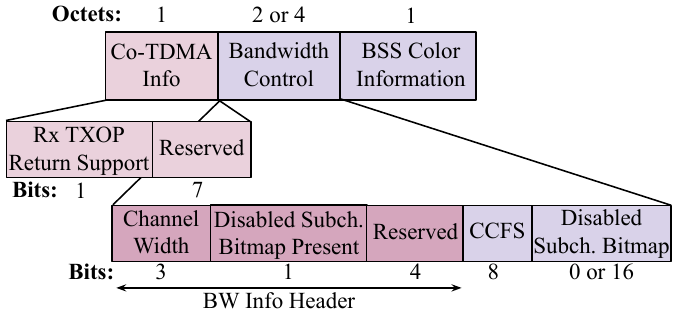}%
\label{fig:cotdma_profile}}
\vfil
\subfloat[]{\includegraphics[width=0.4\textwidth]{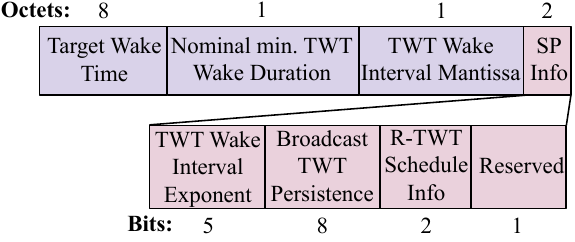}%
\label{fig:cortwt_profile}}
\vfil
\subfloat[]{\includegraphics[width=0.49\textwidth]{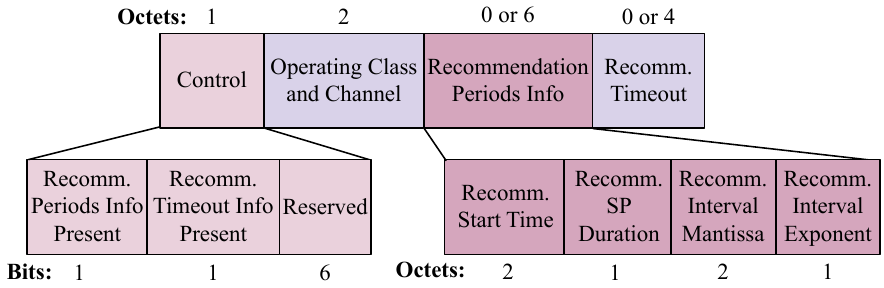}%
\label{fig:cocr_profile}}
\caption{Per-scheme \gls{mapc} parameter sets included in the \gls{mapc} element. (a) \gls{cobf}, (b) \gls{cosr}, (c) \gls{cotdma}, (d) \gls{cortwt}, (e) \gls{cocr}.}
\label{fig:mapc_parameter_set}
\end{figure}

802.11 devices include Information Elements (IEs) in the body of different types of frames (e.g., management, control, action) to inform other nodes about their capabilities and status. The 11bn defines a new element called \texttt{\gls{mapc} Element} (illustrated in Fig.~\ref{fig:mapc_element}), which has been introduced to inform about the \gls{mapc} capabilities and thus enabling the discovery (cf. Section~\ref{sec:mapc_discovery}) and subsequent negotiation (cf. Section~\ref{sec:mapc_negotiation}) of coordination agreements. Inside the \texttt{\gls{mapc} Element}, we find the following new fields:
\begin{itemize}
    \item \texttt{\gls{mapc} control} field (8 bits) includes \texttt{\gls{ap} ID Present} (1 bit set to 1 if the \gls{ap} identifier is present in the \texttt{\gls{mapc} Common Info} field, and set to 0 otherwise) and Reserved (7 bits).
    \item The \texttt{\gls{mapc} Common Info} field (variable size) carries information that is common to all the \gls{mapc} schemes, where:
    \begin{itemize}
        \item The \texttt{\gls{mapc} Common Info Length} field (8 bits) indicates the number of octets in the \texttt{\gls{mapc} Common Info} field, including one octet for the \texttt{\gls{mapc} Common Info Length} field.
        \item \texttt{\gls{mapc} Capabilities} (16 bits) includes a bit to indicate support for each \gls{mapc} feature. In addition, it is indicated whether the \gls{ap} supports \gls{mapc} security and the capability to reply to Trigger frames.
        \item \texttt{\gls{mapc} Parameters} (16 bits) includes one bit for each supported \gls{mapc} feature to indicate whether an agreement with another \gls{ap} has already been established or not. In addition, a \texttt{Protected Negotiations Required} field is used to indicate the need for protected \gls{mapc} agreement negotiations (executed through \emph{Protected \gls{mapc} Negotiation Request} and \emph{Protected \gls{mapc} Negotiation Response} messages).
        \item \texttt{\gls{ap} ID} (0 or 16 bits) is the identifier assigned to the other coordinated \gls{ap}.
    \end{itemize}
    \item The \texttt{\gls{mapc} Schemes Info} field (variable size) carries information specific to one or more \gls{mapc} schemes. This information is presented in the form of subelements inside the \texttt{\gls{mapc} element} (Fig.~\ref{fig:mapc_element}, in purple), which can be used to carry the necessary details of each scheme, but also to inform about other aspects (e.g., security profile for the \gls{mapc} \gls{pasn} authentication described in Section~\ref{sec:mapc_negotiation}). The subelement type is informed using the \texttt{Subelement ID} (from 0 to 255). Focusing on the \texttt{Per-Scheme Profile subelements} (\texttt{Subelement ID} $= 0$), which define the different coordinated transmission schemes. We find the following fields:
    \begin{itemize}
        \item \texttt{\gls{mapc} Scheme Control} (8 bits): Includes the \gls{mapc} scheme type (4 bits indicating \gls{cobf}, \gls{cosr}, \gls{cotdma}, \gls{cortwt}, \gls{cocr}, or 11 reserved values) and 4 reserved bits.
        \item \texttt{\gls{mapc} Scheme Parameter Set} (variable size): Includes specific parameters for the indicated \gls{mapc} scheme, so its shape varies accordingly.
        \begin{itemize}
            \item \texttt{\gls{cobf} profile} (Fig.~\ref{fig:cobf_profile}): Includes the total number of \gls{obss} sounding reports supported by the \gls{ap} (2 bits), the number of joint sounding reports supported by the \gls{ap} for its peer (2 bits), support or not for \gls{gi} of $2\times$ \gls{ltf} + $0.8 \mu$s, and 3 reserved bits.
            \item \texttt{\gls{cosr} profile} (Fig.~\ref{fig:cosr_profile}): Includes the minimum acceptable transmit power by the coordinated \gls{ap} to participate in a \gls{cosr} transmission (6 bits) and 2 more reserved bits.
            \item \texttt{\gls{cotdma} profile} (Fig.~\ref{fig:cotdma_profile}): Includes the \texttt{\gls{cotdma} Info} (8 bits), the \texttt{Bandwidth Control} (16 or 32 bits, with information about the \gls{bss} bandwidth), and the \texttt{\gls{bss} Color Information} (8 bits). 
            \item \texttt{\gls{cortwt} profile} (Fig.~\ref{fig:cortwt_profile}): Includes the necessary information for handling coordinated \glspl{sp} (e.g., \gls{sp} start time, \gls{sp} duration). In particular, it contains the \texttt{Target Wake Time} (8 octets), the \texttt{Nominal Minimum \gls{twt} Wake Duration} (1 octet), the \texttt{\gls{twt} Wake Interval Mantissa} (1 octet), and per-\gls{sp} information. 
            \item \texttt{\gls{cocr} profile} (Fig.~\ref{fig:cocr_profile}): A \texttt{Control} field (1 octet) informs about the presence of \texttt{Recommendation Periods Info} and \texttt{Recommendation Timeout Info}, whose specific details are optionally included in the \texttt{Recommendation Periods Info} (0 or 6 octets). Other relevant parameters are the \texttt{Operating Class and Channel} (2 octets) and the \texttt{Recommendation Timeout} (0 or 4 octets).
        \end{itemize}
        \item \texttt{\gls{mapc} Scheme Request Set} (variable size): This field is used to manage coordination agreements between \glspl{ap} (establishment, update, or teardown). Depending on the type of message and the selected \gls{mapc} scheme, the content of the \texttt{\gls{mapc} Scheme Request Set} varies. In particular, it contains an \texttt{\gls{mapc} Request Control} (8 bits), an \texttt{\gls{mapc} Per-Scheme Info} (8 bits), and an \texttt{\gls{mapc} Request Parameter Set} (variable size). The latter includes specific parameters for request messages, and its content also varies depending on the \gls{mapc} scheme (e.g., for \gls{cortwt}, it includes the identifier of the \gls{twt} schedule).
    \end{itemize}
\end{itemize}

\subsection{MAPC Messages}

\begin{figure}[ht]
\centering
\subfloat[]{\includegraphics[width=.7\columnwidth]{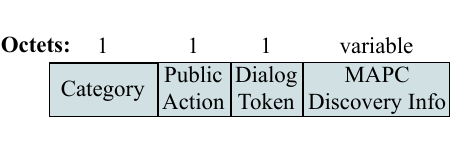}  \label{fig:mapc_discovery_request_response}}
\vfil
\subfloat[]{\includegraphics[width=0.7\columnwidth]{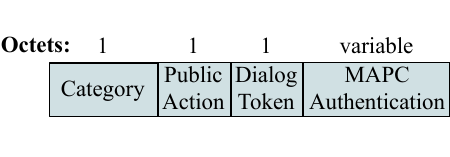}%
\label{fig:mapc_authentication}}
\vfil
\subfloat[]{\includegraphics[width=0.8\columnwidth]{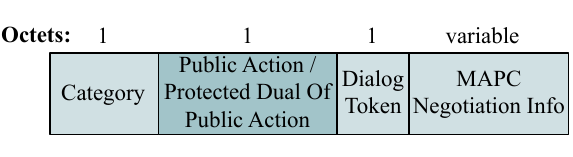}%
\label{fig:mapc_negotiation_request}}
\vfil
\subfloat[]{\includegraphics[width=0.9\columnwidth]{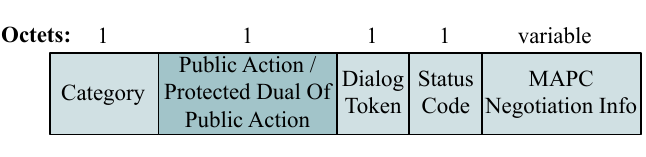}%
\label{fig:mapc_negotiation_response}}
\caption{\gls{mapc} messages. (a) \gls{mapc} Discovery Request/Response, (b) \gls{mapc} Authentication, (c) \gls{mapc} Negotiation Request, (d) \gls{mapc} Negotiation Response.}
\label{fig:mapc_messages}
\end{figure}

To discover \gls{mapc} functionalities from other \gls{ap} and establish negotiation agreements, 11bn introduces new types of \texttt{Public Action} frames:
\begin{itemize}
    \item \texttt{\gls{mapc} Discovery Request} (Fig.~\ref{fig:mapc_discovery_request_response}): Used to advertise the \gls{mapc} capabilities of an \gls{ap} and to request the capabilities of other \glspl{ap}. It includes the \texttt{Category} (8 bits, to indicate it is a public action frame), the \texttt{Public Action Value} (8 bits, set to 66 in this case), and the \texttt{\gls{mapc} Discovery Info} (which contains the \gls{mapc} element described in Appendix~\ref{app:mapc_element}) fields.
    \item \texttt{\gls{mapc} Discovery Response} (Fig.~\ref{fig:mapc_discovery_request_response}): Used by an \gls{ap} to reply to an \texttt{\gls{mapc} Discovery Request} and announce its capabilities. The same frame as for the request (but indicating 67 in the Public Action Value field) is used.
    \item \texttt{\gls{mapc} Authentication} (Fig.~\ref{fig:mapc_authentication}): Used to carry the three-step \gls{mapc} \gls{pasn} authentication for negotiating \gls{mapc} agreements. In this case, the \texttt{\gls{mapc} Authentication} field is used to act as \textit{Initiation/Key Request} (includes the \gls{rsne}, \gls{pasn} parameters element, and conditionally the \gls{rsnxe}), \textit{Response/Key Confirmation} (includes the \gls{mic} element, \gls{pasn}, and \gls{rsne}/\gls{rsnxe}), and \textit{Final Confirmation} (includes the \gls{mic} element and final \gls{pasn} parameters).
    \item \texttt{\gls{mapc} Negotiation Request} (Fig.~\ref{fig:mapc_negotiation_request}): Used to initiate a negotiation for a specific scheme. Depending on the need or not for authentication (known after the discovery), this frame can be either a \texttt{Public Action} or a \texttt{Protected Dual of Public Action frame}. In both cases, it includes a \texttt{\gls{mapc} Negotiation Info} field (with the \gls{mapc} element) to inform about the proposed configuration.
    \item \texttt{\gls{mapc} Negotiation Response} (Fig.~\ref{fig:mapc_negotiation_response}): Used to acknowledge and respond to a \texttt{\gls{mapc} Negotiation Request}. Apart from reusing the same fields from the request, it also includes a status code.
\end{itemize}

\bibliographystyle{IEEEtran}
\bibliography{references}

 
\begin{IEEEbiography}[{\includegraphics[width=1in,height=1.25in,clip,keepaspectratio]{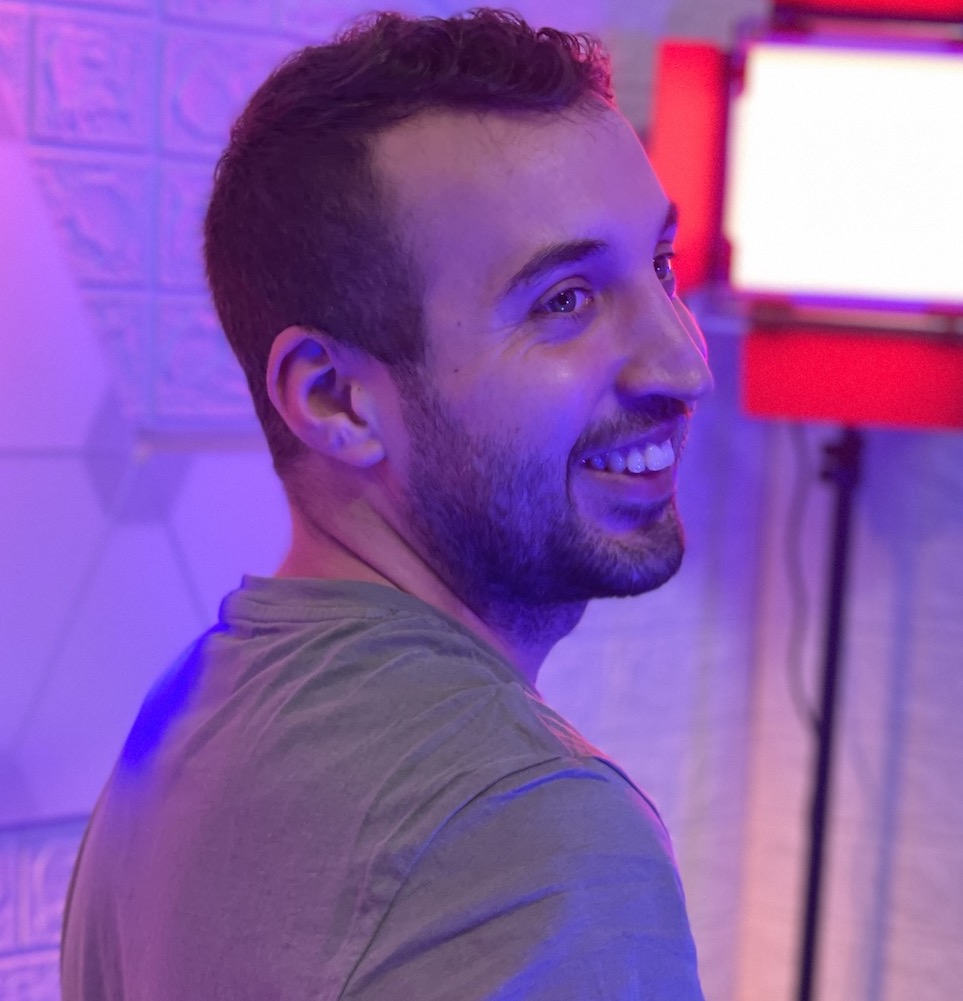}}]{Francesc Wilhelmi}
(Member, IEEE) holds a Ph.D. in information and communication technologies (2020) from the Universitat Pompeu Fabra (UPF). He also holds a B.Sc. degree in telematics engineering (2015) and an M.Sc. degree in intelligent and interactive systems (2016), also from the UPF. He is currently a Tenure Track Professor at Universitat Pompeu Fabra (UPF). Before, he was with Nokia Bell Labs and the Catalan Technology Center for Telecommunications (CTTC). In 2024, he received the IEEE ComSoc Best Young Professional Award.
\end{IEEEbiography}

\begin{IEEEbiography}[{\includegraphics[width=1in,height=1.5in,clip,keepaspectratio]{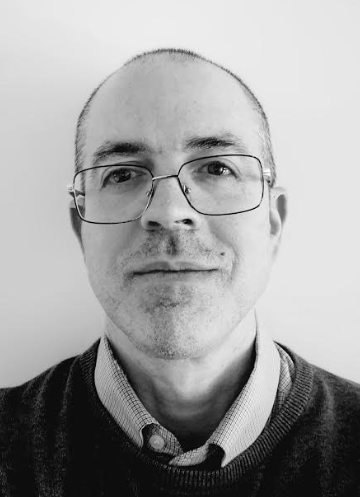}}]{Boris Bellalta} (Senior Member, IEEE)
is a Professor at Universitat Pompeu Fabra (UPF), where he leads the Wireless Networking group. His research interests lie in the area of wireless networks and performance evaluation, with a particular emphasis on Wi-Fi technologies and machine learning-based adaptive systems.  
\end{IEEEbiography}

\begin{IEEEbiography}[{\includegraphics[width=1in,height=1.25in,clip,keepaspectratio]{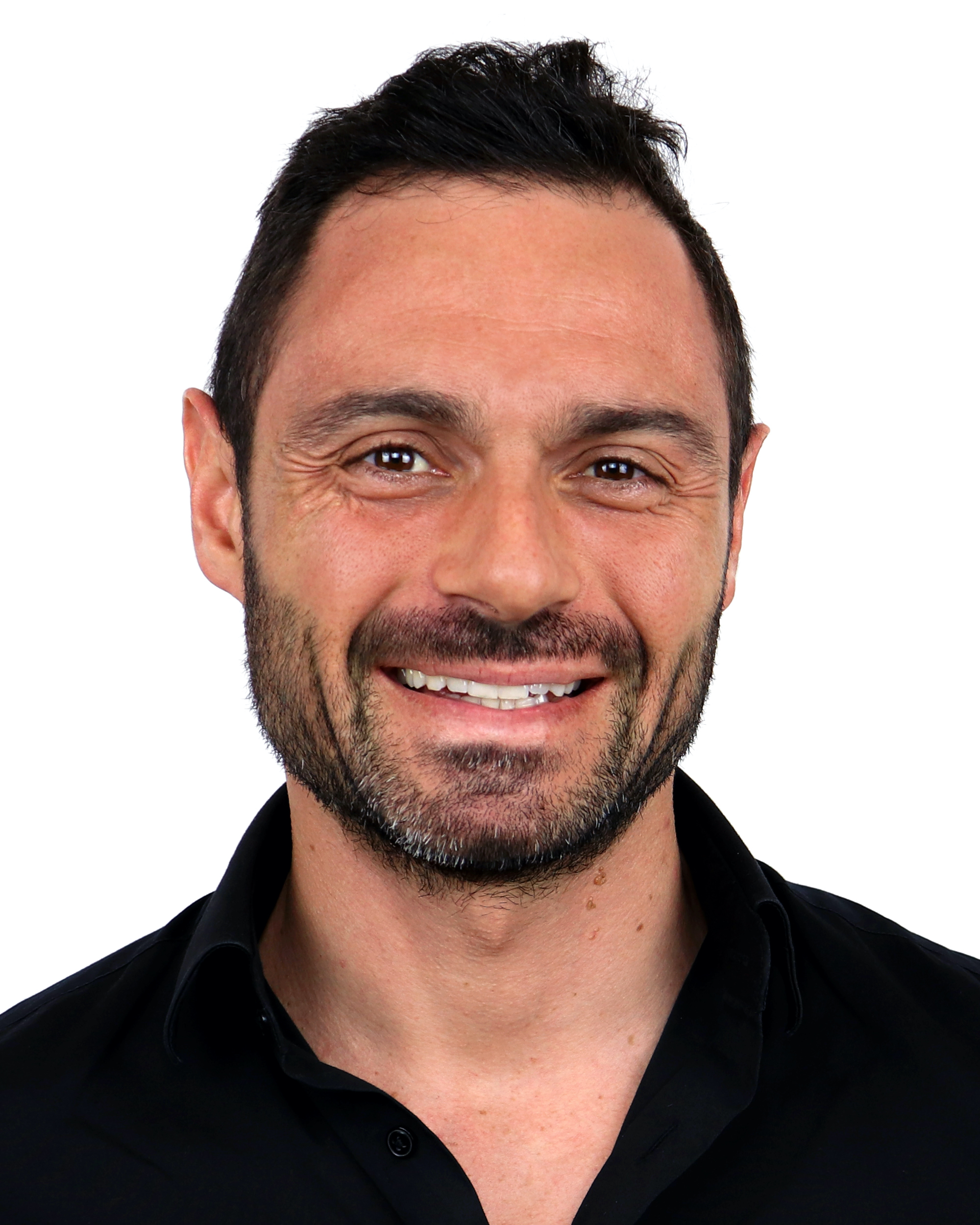}}]
{Giovanni Geraci} (Senior Member, IEEE)
is a Team Leader at Nokia Technology Standards and an Associate Professor at Universitat Pompeu Fabra, Spain. Previously, he was a Principal Research Scientist at Telefónica and a Research Scientist at Nokia Bell Labs. He served as an IEEE Distinguished Lecturer and created the IEEE course “Fundamentals of Wi-Fi”. He received the IEEE Communications Theory Early Achievement Award (2023), the IEEE Communications Society EMEA Outstanding Young Researcher Award (2018), and Best Paper Awards at IEEE ICC (2025), IEEE GLOBECOM (2022), and IEEE PIMRC (2019).
\end{IEEEbiography}

\begin{IEEEbiography}[{\includegraphics[width=1in,height=1.25in,clip,keepaspectratio]{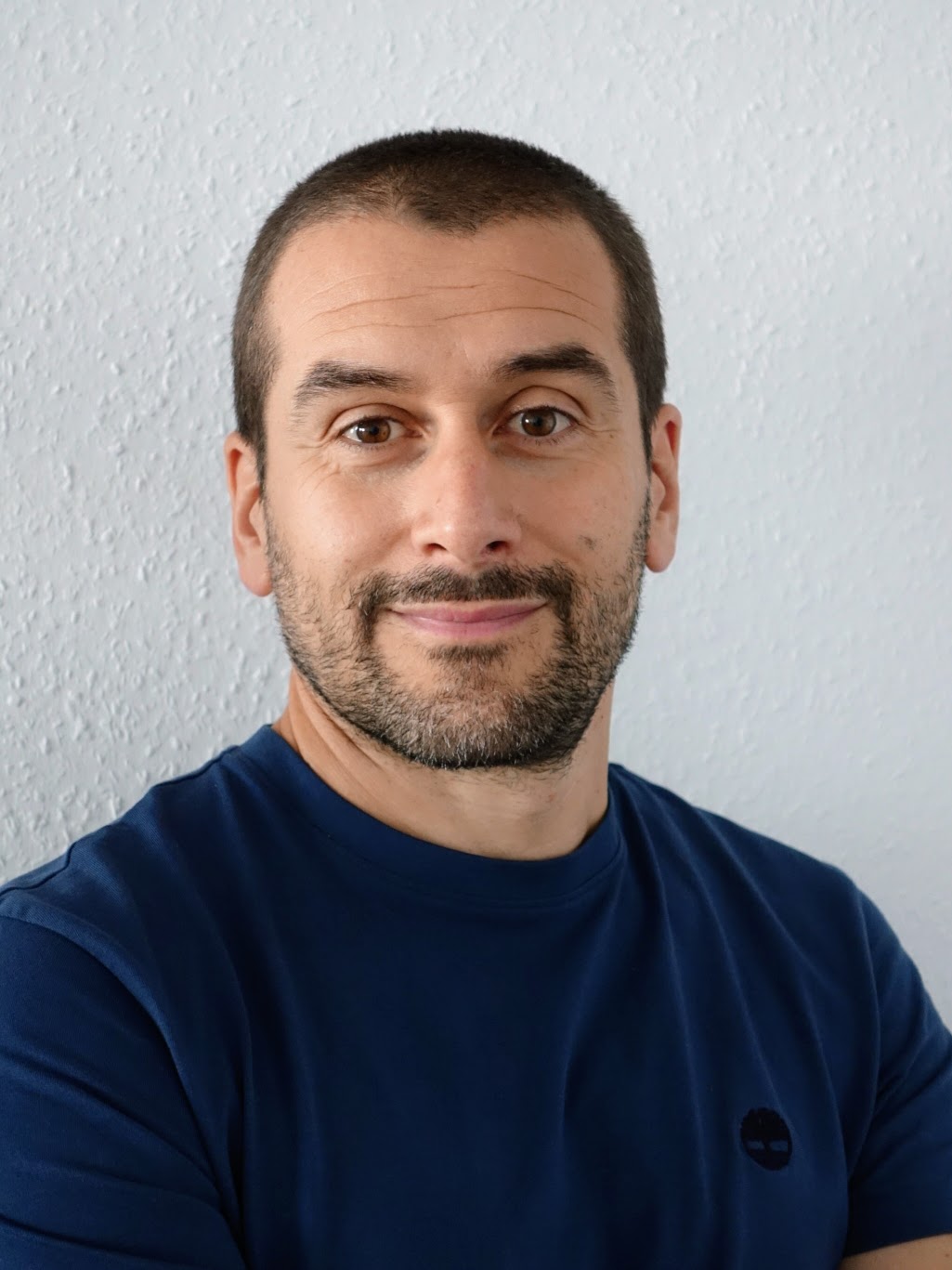}}]
{Lorenzo Galati Giordano} (Senior Member, IEEE)
is a Distinguished Member of Technical Staff (DMTS) at Nokia Bell Labs, where he has been contributing since 2015. With nearly two decades of academic and industrial experience in communication systems, protocols, and standards, Lorenzo has made significant research contributions in the fields of massive MIMO and radio systems operating in the unlicensed spectrum. His work has led to the co-authorship of numerous commercial patents and publications in esteemed books, IEEE journals and conferences, as well as contributions to IEEE 802.11 standards. Currently, Lorenzo is leading research initiatives focused on innovative techniques and AI/ML-based optimizations for reliable, low-latency communications in Wi-Fi. Prior joining Nokia Bell Labs, he served as an R\&D System Engineer at Azcom Technology from 2010 to 2014. Lorenzo holds a Ph.D. from Politecnico di Milano in Italy and a post-graduate master's degree in Innovation Management from IlSole24Ore Business School, Italy.
\end{IEEEbiography}

\begin{IEEEbiography}[{\includegraphics[width=1in,height=1.25in,clip,keepaspectratio]{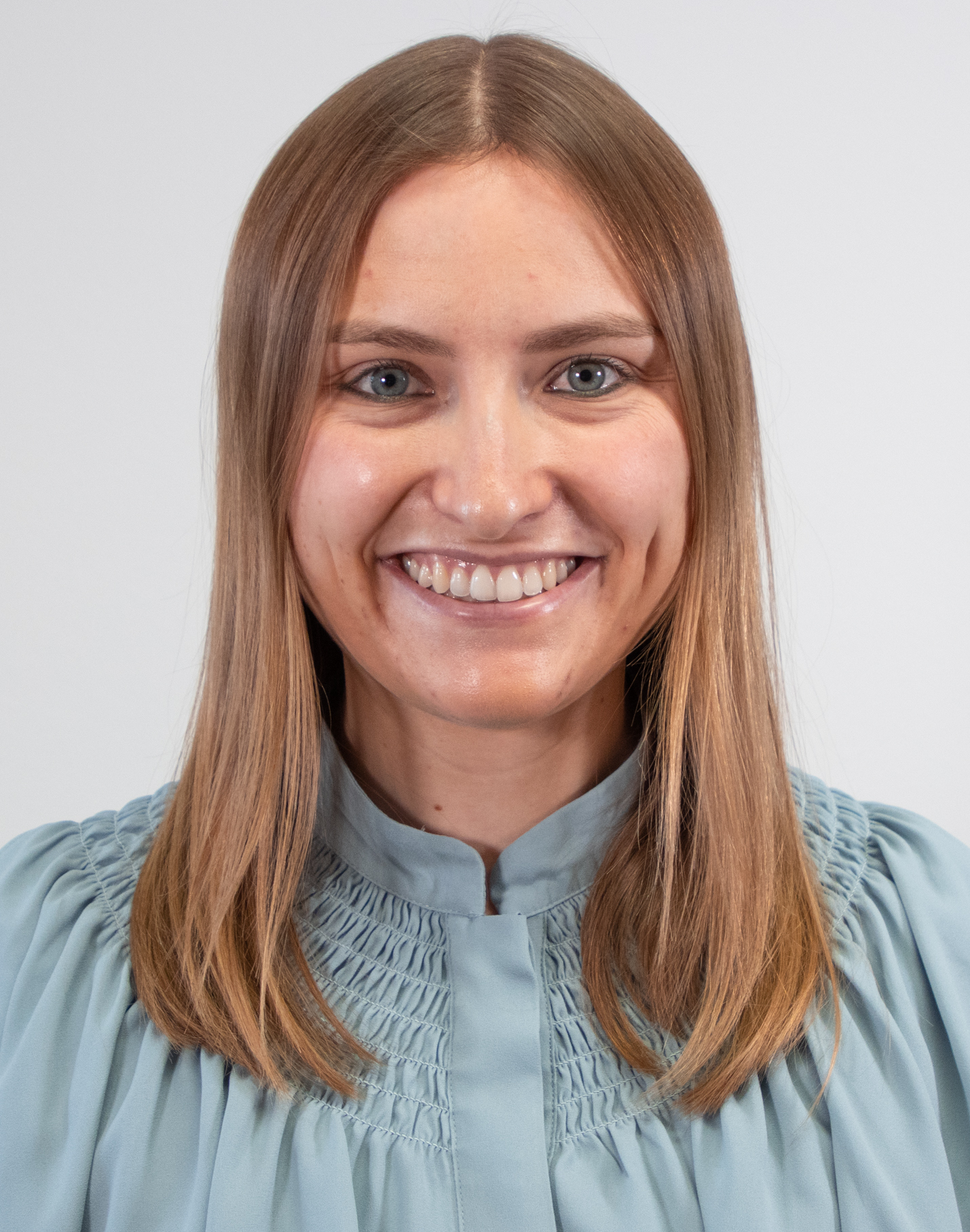}}]
{Francesca Meneghello} (Member, IEEE)
is a Principal Research Scientist at the Institute for Networked Intelligent Systems (INSI) at Northeastern University, USA. She worked as an Assistant Professor at the Department of Information Engineering at the University of Padova, Italy. Her work focuses on wireless communications and machine learning, contributing to defining and developing next-generation data-driven wireless technologies that can adapt to the context. She served as TPC co-chair of the IEEE INFOCOM DeepWireless 2025 Workshop and guest editor of npj Wireless Technology. She has received a Best Paper Award at IEEE INFOCOM 2025 and IEEE INFOCOM 2026. She is a 2023 Fulbright-Schuman alumna and is a recipient of an MSCA Global Postdoctoral Fellowship funded by the European Commission under the Horizon Europe scheme (2026–2029).
\end{IEEEbiography}

\begin{IEEEbiography}[{\includegraphics[width=1in,height=1.25in,clip,keepaspectratio]{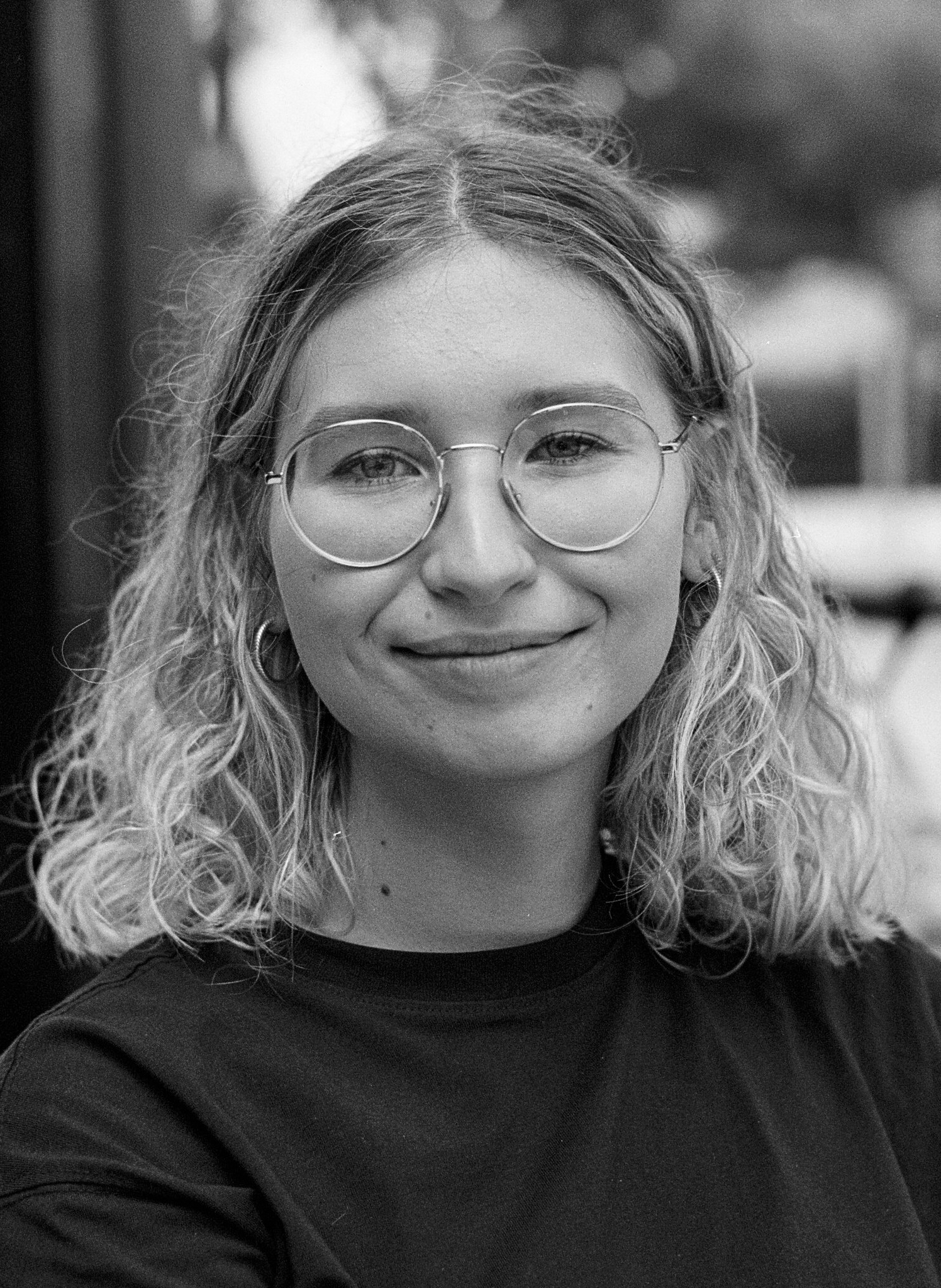}}]{Aleksandra Kijanka}
received her M.Sc. in Information and Communication Technologies (ICT) from AGH University of Krakow in 2024. She completed her master’s thesis work at Universitat Politècnica de València (UPV) under Prof. Narcís Cardona, focusing on material characterization for channel modeling. Following a Wi-Fi research project at UPV, she began her Ph.D. at Universitat Pompeu Fabra (UPF) under the supervision of Prof. Boris Bellalta. Her doctoral work explores machine learning (ML) applications to optimize Wi-Fi performance. Additionally, she gained industry experience at Nokia Krakow (2022–2023), contributing to 5G base station software development.
\end{IEEEbiography}

\begin{IEEEbiography}[{\includegraphics[width=1in,height=1.25in,clip,keepaspectratio]{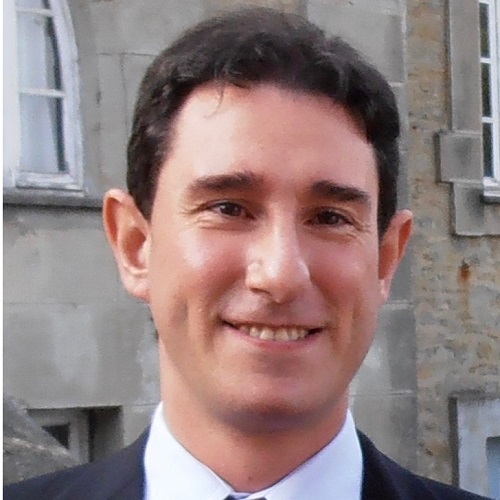}}]
{Iñaki Val} (Senior Member, IEEE)
is a Principal Engineer of Standards Engineering at MaxLinear. He was previously with Ikerlan Research Institute as Team Leader of the communication systems department. He actively collaborates with IEEE and engineering departments in the definition of next-generation Wi-Fi technologies. His technical background includes more than 20 years of experience in embedded electronic systems and in the design, implementation, and validation of real-time and reliable wireless communication solutions for academic research and industrial customers.
\end{IEEEbiography}

\begin{IEEEbiography}[{\includegraphics[width=1in,height=1.25in,clip,keepaspectratio]{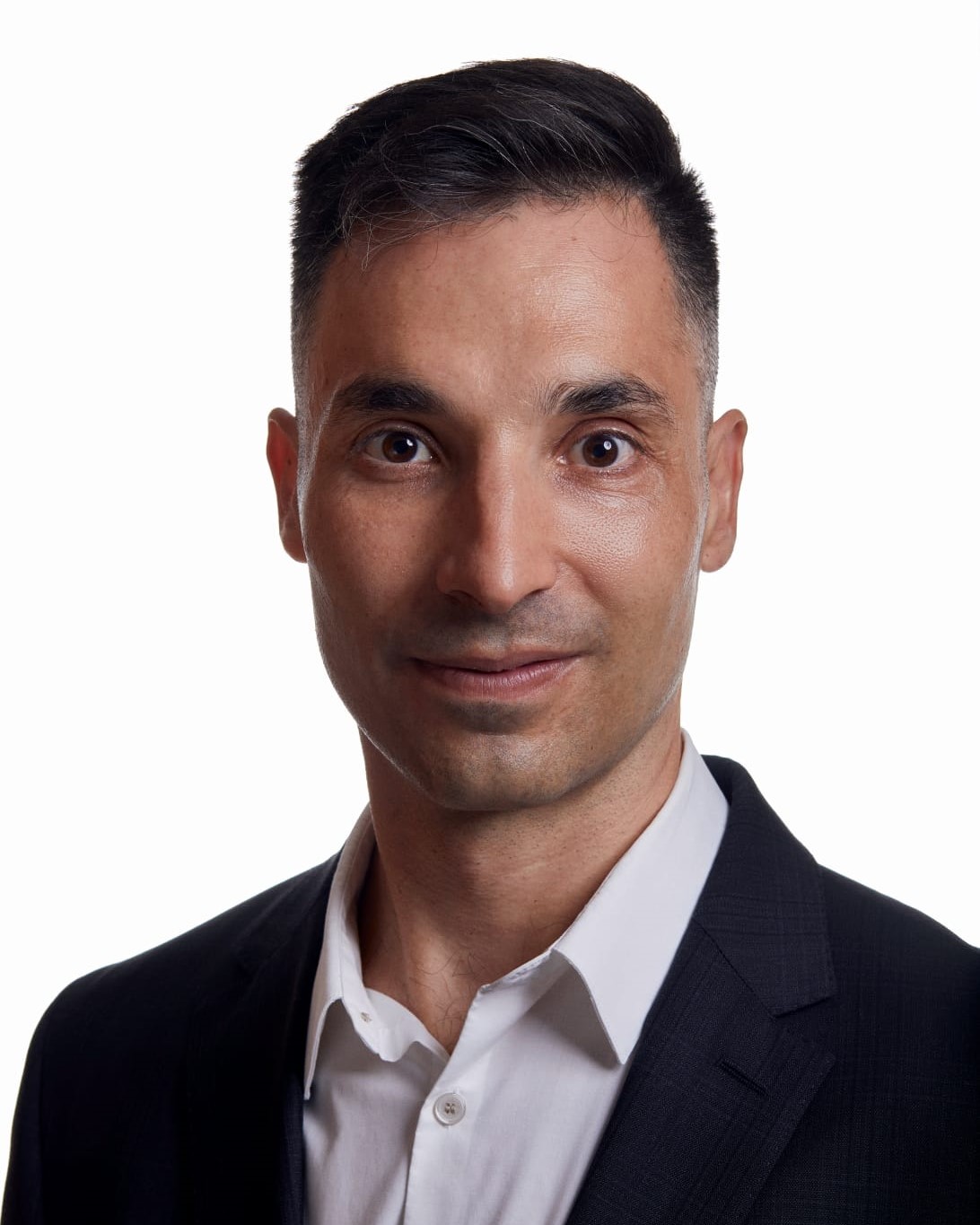}}]
{David López-Pérez} (Senior Member, IEEE) is Associate Professor at Universitat Politècnica de València. Before joining this position, David served as an Expert and Technical Leader at Huawei Technologies in Paris and held the title of Distinguished Member of Staff at Nokia Bell Labs in Dublin. Throughout his career, David has focused on the study of both cellular and Wi-Fi networks, with primary research interests in network performance analysis, network planning and optimization, heterogeneous networks,  non-terrestrial networks, green networking, machine learning, and the development of new technologies and features.
\end{IEEEbiography}

\end{document}